\documentclass[acmsmall,screen,normalem]{acmart}
\setcopyright{none}
\renewcommand\footnotetextcopyrightpermission[1]{}
\usepackage{graphicx} 
\usepackage{booktabs}
\usepackage{multirow}
\usepackage{ulem}
\usepackage{colortbl}
\usepackage{subcaption}
\usepackage{multicol}
\usepackage{color}
\usepackage{listings, listings-rust}
\usepackage{float}
\usepackage{hyperref}
\usepackage{bussproofs}
\usepackage{makecell}
\usepackage{algorithm}
\usepackage{algpseudocode}
\usepackage{tcolorbox}
\usepackage{hyperref}
\usepackage{enumitem}

\newtcbox{\mybox}{on line,
  arc=2pt, outer arc=2pt,    
  colback=gray!20,           
  colframe=gray!20,          
  boxsep=0pt, left=3pt, right=3pt, top=2pt, bottom=2pt, 
  boxrule=0pt, bottomrule=0pt, toprule=0pt,
  fontupper=\sffamily\bfseries\small 
}

\newcommand{\solution}[1]{\mybox{#1}~}

\usepackage{amsmath, amssymb, amsthm}
\newtheorem{definition}{Definition}

\newtheoremstyle{plainupright} 
  {}{}                          
  {\normalfont}    
  {}           
  {\normalfont}        
  {.}                     
  { }                         
  {\thmname{#1}~\thmnumber{#2}} 

\theoremstyle{plainupright}
\newtheorem{infrule}{Inference Rule}

\def\BibTeX{{\rm B\kern-.05em{\sc i\kern-.025em b}\kern-.08em
    T\kern-.1667em\lower.7ex\hbox{E}\kern-.125emX}}

\begin{document}
\title{Annotating and Auditing the Safety Properties of Unsafe Rust}

\author{Zihao Rao}
\email{zhrao25@m.fudan.edu.cn}
\affiliation{%
  \institution{Fudan University}
  \country{China}
}

\author{Jiping Zhou}
\email{zjp1137967378@gmail.com}
\affiliation{%
  \institution{Independent Researcher}
  \country{China}
}

\author{Hongliang Tian}
\email{tate.thl@antgroup.com}
\affiliation{%
  \institution{Ant Group}
  \country{China}
}

\author{Xin Wang}
\email{xinw@fudan.edu.cn}
\affiliation{%
  \institution{Fudan University}
  \country{China}
}

\author{Hui Xu}
\authornote{Corresponding author}
\email{xuh@fudan.edu.cn}
\affiliation{%
  \institution{Fudan University}
  \country{China}
}

\begin{abstract}
In Rust, unsafe code is the sole source of potential undefined behaviors.
To avoid misuse, Rust developers should clarify the safety properties for each unsafe API.
However, the community currently lacks a key standard for safety documentation: existing safety comments in the source code and safety documentation can be ad hoc and incomplete.
This paper presents a tag-centric methodology for auditing the consistency and completeness of safety documentation.
We first derive a taxonomy of \textit{Safety Tags} to formalize natural-language requirements.
Second, because API soundness frequently relies on struct invariants, we propose a set of empirical rules to systematically audit the structural consistency of safety documentation.
We implemented this methodology in \textit{safety-tool}, a static linter that automatically enforces structural consistency between local safety annotations and callee requirements.
Our approach was applied to the Rust standard library, fixing documentation issues on 27 APIs with 61 safety tags and identifying safety tags that are applicable to 96.1\% of the public unsafe APIs in libstd. Furthermore, we have formalized the tagging idea through a Rust RFC to the wider community. We believe that the approach establishes a standardized practice of safety documentation and helps significantly reduce safety perils.
\end{abstract}

\maketitle

\section{Introduction}\label{introduction}
Rust is a systems programming language with strong memory-safety guarantees. Nevertheless, low-level operations such as raw-pointer dereference and unchecked memory manipulation remain unavoidable in systems software, and they continue to be an important source of safety bugs~\cite{peng2024framekernel,sharma2024rust,li2024empirical}. 
Rust prevents undefined behavior as long as programmers remain within Safe Rust, where ownership, borrowing, and lifetime rules are enforced statically~\cite{xu2021memory}. 
By design, the Rust compiler withholds its safety guarantees within unsafe blocks; therefore, the burden of upholding safety invariants shifts primarily to the developer when using unsafe code~\cite{working-with-unsafe}. While guidelines such as the Rustonomicon~\cite{working-with-unsafe} exist, achieving sound encapsulation is still prone to errors, as it requires that developers carefully reason about subtle safety invariants across usage boundaries. As a result, unsoundness often lurks in functions that contain internal unsafe code, making it a characteristic Rust-specific bug class.~\cite{xu2021memory,qin2020understanding}. Such issues could undermine Rust's memory safety feature if they are not properly addressed.
Therefore, the memory safety of Rust ultimately depends on the correct usage of unsafe code.

In this setting, safety documentation plays a critical role: each unsafe API should clearly specify its safety requirements in its documentation.
Auditing such documentation is therefore essential for identifying missing or misstated requirements. In practice, however, both establishing correct safety documentation and maintaining it over time remain difficult for three main reasons.
First, this audit is non-trivial because the relevant requirements are often not presented in a formal, uniform, and directly checkable form. Instead, they may be scattered across comments, cross-references, and side-effect warnings~\cite{cui2024unsafe}, and similar requirements may be expressed differently across APIs. Auditors therefore often need to reconstruct the intended contract before they can assess whether the documentation is correct and complete.
Second, the safety requirements that should be documented for an API are not always determined locally. For example, the soundness of a safe method may depend on struct invariants established by constructors, and an unsafe constructor may therefore need to document the conditions required to establish or preserve those invariants. Auditing the documentation thus often requires tracing safety obligations across multiple functions and data structures rather than examining each API in isolation.
Third, even after an audit has been completed, software evolution can introduce new safety requirements or alter existing ones, creating hidden risks for downstream users. When the safety contract of an internal unsafe API changes, its downstream callers may also adapt accordingly. Because Rust's type signatures do not encode such documentation-level obligations, and the compiler provides no warning when they change, downstream code may remain unaware of newly introduced requirements and omit the corresponding checks, ultimately leading to undefined behavior.

To address these challenges, we propose an auditing methodology that combines human semantic judgment with automated structural analysis\footnote{\url{https://github.com/safer-rust/safety-tags}}. The methodology comprises three core components: (1) a taxonomy of \textit{safety tags} for representing safety requirements; (2) structural dependency extraction for APIs that should be audited together; and (3) \texttt{safety-tool} for checking whether the resulting documentation remains consistent as the code evolves. Together, these components define a four-phase workflow: deriving safety tags, identifying structurally related APIs, auditing their documentation with explicit rules, and continuously checking documentation consistency.

We begin by introducing an annotation language that translates informal safety comments in the Rust standard library into machine-checkable \textit{safety tags}. While contract-based approaches~\cite{astrauskas2022prusti, lattuada2023verus, lattuada2024verus} offer one path toward formalizing safety obligations, they are typically designed for automated proving rather than lightweight documentation. In our framework, each safety requirement is defined as a parameterized tag, \texttt{tag(args) $\mapsto$ expression(args)} (Table~\ref{tab:sp}), together with an explicit contractual role such as precondition, hazardous state, or optional condition. Based on Rust's UB documentation~\cite{ub, working-with-unsafe}, we derive 21 primitive safety tags across five categories: \textit{layout}, \textit{pointer validity}, \textit{content}, \textit{alias}, and \textit{contextual}. These tags cover 96.1\% of public unsafe APIs in the Rust standard library, excluding platform-specific intrinsics that lack generalizable requirements.

We then extract structural dependencies to assemble \textit{Audit Units}, each centered on a target API and its associated constructors. This allows auditors to track how safety tags are handled, delegated, and maintained across call chains and object construction. Within each audit unit, we apply empirical audit rules to check documentation consistency at three layers: immediate call-site logic, invariant initialization, and state exposure.

To maintain audit validity as the software evolves, we developed \texttt{safety-tool}, an IDE-integrated static linter. This tool can expand structured safety tags into human-readable documentation and check consistency between caller annotations and callee requirements. Furthermore, we have submitted a Rust RFC~\cite{rust_rfc_3842} to push these annotations toward ecosystem-wide standardization and tool support.

We evaluate the methodology through an empirical audit of the Rust standard library. We translate informal safety comments into formal tags, extract structural dependencies to infer the safety obligations in the source code, and compare these inferred obligations with the existing documentation to identify inconsistencies. This process uncovers 27 APIs with 61 defective safety tags. To date, the official Rust project has already confirmed and merged our fixes for all of these tags.

In summary, this paper makes the following contributions.
\begin{itemize}[leftmargin=*]
    \item We derive a taxonomy of 21 safety tags from the Rust standard library and propose empirical audit rules that use these tags to check the consistency and completeness of API safety documentation.
    \item We implement \textit{safety-tool}, a static linter that enforces tag consistency between caller annotations and callee requirements, and submit a formal Rust RFC to integrate the system into the Rust ecosystem.
    \item We audit the Rust standard library, identify documentation defects in 27 APIs involving 61 safety tags, and contribute fixes that have all been merged into the official Rust project, reducing the potential for unsafe API misuse in practice.
\end{itemize}

\section{Safety Promise and Unsafety Perils of Rust}
\subsection{Memory Safety of Rust}

As a system language, Rust facilitates fine-grained memory management and efficiency with raw pointers and low-level controls. Their full elimination is impossible by design in \texttt{unsafe} blocks, which allow developers to bypass the compiler's checks. Moreover, the undecidability of precise pointer analysis makes mechanical verification of unsafe code extremely difficult~\cite{hind2001pointer, landi1992undecidability}. Rust’s safety promise is that ``\textit{the program should not exhibit undefined behaviors if developers do not use unsafe code}''~\cite{xu2021memory}. This safety model relies on the interplay between two modes: Safe Rust and Unsafe Rust.

\subsubsection{Safe and Unsafe Rust} Safe Rust achieves the promise of safety through several novel features, including ownership and lifetimes. In safe Rust, each value has an owner, and ownership can be borrowed either mutably or immutably. The compiler mandates \textit{exclusive mutability}: if a value has a mutable reference, it cannot be aliased. It can avoid the tricky pointer analysis problems but at the cost of flexibility. For instance, implementing a doubly linked list in safe Rust is challenging because it requires multiple mutable references to the same node.

To regain flexibility, Rust permits developers to bypass safety rules via \texttt{unsafe} code. Common unsafe operations include dereferencing raw pointers, calling unsafe functions, accessing mutable static variables, and accessing union fields. Notably, implementing an unsafe trait is different because it signifies that the implementation must maintain extra invariants~\cite{unsafe-rust}. Since the compiler does not guarantee memory safety in unsafe blocks, developers bear the responsibility for potential risks. Unsafe Rust thus complements safe Rust by enabling low-level control and interoperability in systems contexts such as embedded systems~\cite{sharma2024rust} and OS development~\cite{peng2024framekernel, peng2025asterinas, tianasterinas}.

\subsubsection{Unsafe Code Encapsulation}\label{sec:encapsulation}
As unsafe code is often unavoidable, encapsulating it is essential to limit risk propagation. The Rust community advocates hiding unsafe code behind \textit{safe abstractions}~\cite{working-with-unsafe,peng2024framekernel,astrauskas2020programmers}, a practice known as \textit{interior unsafe}~\cite{qin2020understanding, qin2024understanding} where a safe function contains unsafe code and manages all potential undefined behaviors internally. For example, in Figure~\ref{fig:challenges}, the method \texttt{St2::get()} is safe despite calling the unsafe \texttt{slice::from\_raw\_parts()} inside, because \texttt{St2's} unsafe constructor functions must be ensured to uphold safety for themselves and the safe \texttt{St2::get()}. This pattern prevents unsafety from leaking and is widely used in Rust libraries.

\subsubsection{Modularized Safety} Rust provides visibility control to prevent unauthorized access~\cite{visibility}. A data structure or module is considered to be sound if using its public APIs without \texttt{unsafe} code does not cause undefined behavior. Accessing public fields should therefore be conducted with caution, so as not to cause undefined behavior in methods or compromise module safety. Unsafe fields~\cite{unsafe-fields} are an effort to enhance the safety of field access within or across modules.

\subsection{Motivation: Audit Objectives and Challenges}\label{sec:perils}
As previously discussed, Rust allows unsafe blocks for low-level control, making it a focus for system soundness.
While existing research uses automated tools to analyze code logic~\cite{miri, vanhattum2022verifying, lattuada2023verus}, verifying whether documentation matches code requirements still depends on manual auditing. This section defines the objectives of documentation auditing and identifies three challenges in the current workflow.

\subsubsection{Audit Objective}
The objective of documentation auditing is to check whether the safety requirements are correct and complete. An unsafe API is sound if it prevents undefined behavior when the caller meets its safety requirements. This requires checking both the code and the safety documentation. 
The documentation serves to formalize and expose this underlying contract.
Incomplete documentation increases the risk of API misuse.
Listing~\ref{case2} demonstrates a sound example. The soundness of \texttt{create\_non\_null\_ptr\_unchecked} is conditional: it guarantees safety only if the caller strictly adheres to the documented contract that \texttt{ptr} must be non-null.

\begin{lstlisting}[language=Rust,style=boxed, caption={A simple case of unsafe API soundness. The unsafe function propagates the \textit{non-null} requirement to the caller.}, label={case2}]
/// # Safety: `ptr` must be non-null.
pub unsafe fn create_non_null_ptr_unchecked(ptr: *mut u8) -> NonNull<u8> { 
    unsafe { NonNull::new_unchecked(ptr) }
} 
\end{lstlisting}

\subsubsection{Challenges of Standard Audit}\label{sec:challenges}

\begin{figure}[t]
    \centering
    \includegraphics[width=\linewidth]{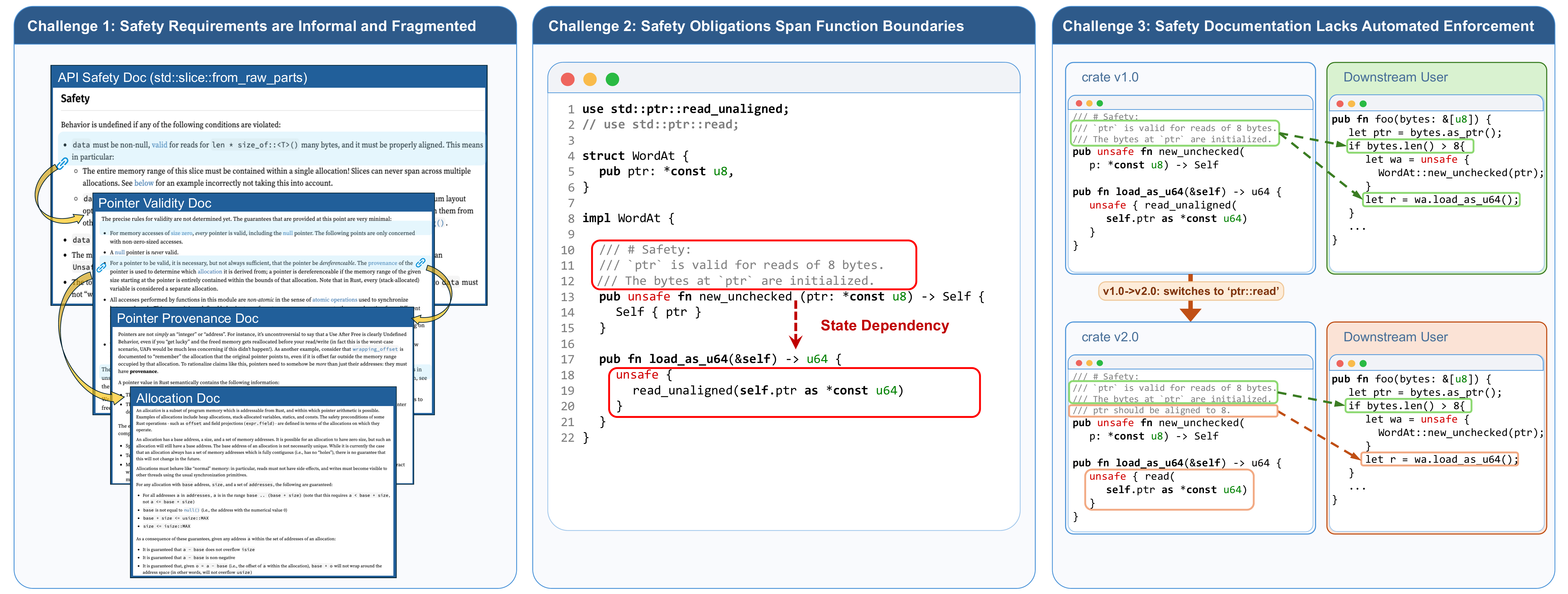}
    \caption{Challenges of standard audit workflow.}
    \label{fig:challenges}
\end{figure}

The standard auditing workflow requires auditors to examine internal unsafe operations within an API to determine the safety requirements. They must then check whether the API satisfies these requirements internally or declares them in its public documentation. However, as shown in Figure~\ref{fig:challenges}, this manual process faces three challenges.

\phantomsection\label{chal:1}\textit{\uline{Challenge 1: Safety Requirements are Informal and Fragmented.}}
The first challenge is that existing safety documentation does not provide an immediate basis for determining whether an API's documented contract is correct and complete (Figure~\ref{fig:challenges}, left). 
This difficulty has two concrete causes.
First, relevant conditions are often distributed across safety comments, cross-references, and other sections. For example, ambiguous phrases such as \textit{valid pointer} in the documentation of \texttt{slice::from\_raw\_parts} require auditors to follow chains of hyperlinks to recover the intended constraints.
Second, safety requirements are not documented in a uniform style~\cite{cui2024unsafe}. Some APIs state preconditions (\textit{e.g.}, non-null requirement in \texttt{NonNull::new\_unchecked} ), while some merely imply constraints through side-effect warnings (\textit{e.g.}, the double-free warning in \texttt{Box::from\_raw} ). This inconsistency makes auditors rely on subjective interpretation to reconstruct the contracts.

\phantomsection\label{chal:2}\textit{\uline{Challenge 2: Safety Obligations Span Function Boundaries.}}
The safety of an API often depends on the internal state of its struct. As shown in Figure~\ref{fig:challenges}, \texttt{WordAt::new\_unchecked(ptr)} only stores a raw pointer, while the later safe method \texttt{load\_as\_u64()} internally calls \texttt{read\_unaligned(self.ptr as *const u64)}. Therefore, the safety documentation of the constructor cannot be written from the local code of \texttt{new\_unchecked} alone. It should also cover the assumptions introduced by \texttt{load\_as\_u64()}, such as that the 8 bytes at \texttt{ptr} are readable and initialized. In other words, when auditing \texttt{new\_unchecked}, auditors have to infer the obligations it should declare from later uses of the stored state, rather than inspect the constructor in isolation. Because these obligations are propagated through shared struct state, function-local inspection or call-graph analysis may not fully recover the safety requirements of a single API.


\phantomsection\label{chal:3}\textit{\uline{Challenge 3: Safety Documentation Lacks Automated Enforcement.}}
The third challenge is that changes in safety documentation are not flagged automatically. As shown in Figure~\ref{fig:challenges}, in crate v1.0, \texttt{load\_as\_u64()} uses \texttt{read\_unaligned}, so the documentation of \texttt{new\_unchecked(ptr)} only needs to state that the 8 bytes at \texttt{ptr} are readable and initialized. In crate v2.0, the API signature remains unchanged, but the internal implementation switches to \texttt{ptr::read}. The constructor documentation must therefore additionally require that \texttt{ptr} be 8-byte aligned. Both versions are reasonable: the change comes from the evolution of the crate's internal unsafe implementation. However, downstream users may still keep the old wrapper. In Figure~\ref{fig:challenges}, \texttt{foo} still checks only \texttt{bytes.len() > 8} and does not enforce alignment. Since the compiler cannot detect such documentation-level contract changes, this kind of mismatch can be easily missed during evolution.
\section{Overview of Tag-Centric Audit Methodology}
\begin{figure}[t]
    \centering
    \includegraphics[width=\linewidth]{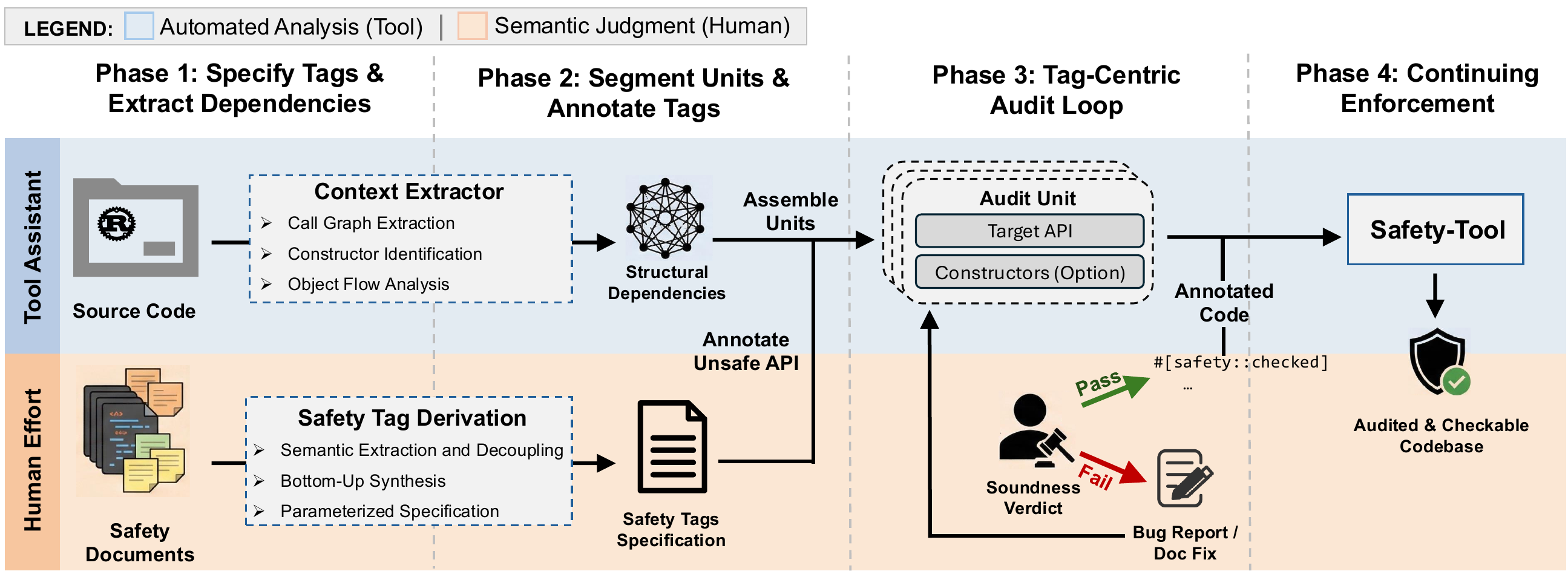}
    \caption{\textbf{General methodology of tag-centric audit}. The methodology yields a safety tag set for the Rust standard library and extracts structural dependencies to assemble audit units, with continuous enforcement provided by \textit{safety-tool}.}
    \label{fig:methodology}
\end{figure}

To address these challenges, we propose a \textit{tag-centric auditing methodology}. In Rust, \texttt{unsafe} blocks are the sole source of undefined behavior (UB)~\cite{ub_promise}. Therefore, the auditing task is to check how the safety obligations are handled. An API must either encapsulate these obligations through internal checks to remain safe, or explicitly annotate them through documentation. Based on this principle, our approach introduces a four-step workflow (Figure~\ref{fig:methodology}) that combines manual semantic analysis with automated dependency tracking. Specifically, it transforms informal documentation into machine-checkable \textit{semantic tags} and traces their propagation across the codebase to verify that all safety obligations are correctly handled.

\textbf{Phase 1: Tag Derivation and Dependency Extraction.} 
This phase prepares the semantic and structural basis for the audit. First, we derive a taxonomy of safety tags from natural-language documentation. This step transforms informal requirements into formal, parameterized tags. Second, we extract structural dependencies from the source code, tracking call graphs and object flows to guide the analysis.

\textbf{Phase 2: Audit Unit Construction.} 
Based on the extracted dependencies, we group related APIs to construct \textit{audit units}. This defines the exact scope of documentation auditing. Concurrently, we annotate unsafe APIs with the derived safety tags, turning abstract properties into concrete audit targets in the source code.

\textbf{Phase 3: The Tag-Centric Auditing.} 
In this phase, auditors examine each \textit{audit unit} using empirical rules. They verify whether the code logic internally satisfies the safety tags or correctly delegates them to the caller's documentation. Inconsistencies or missing documentation are then reported and fixed.

\textbf{Phase 4: Automated Enforcement.} 
To prevent documentation from becoming outdated as software evolves, we introduce \texttt{safety-tool} as a static linter. It automatically verifies the safety tags between callers and callees, ensuring continuous alignment between upstream documentation and downstream code implementations.

\paragraph{Roadmap} The remainder of this paper details the components of this methodology, using the Rust standard library as a comprehensive case study. Section \ref{sec:annotation} defines the taxonomy of safety tags derived from the standard library (challenge \hyperref[chal:1]{1}).
Section \ref{sec:method} elaborates on the audit unit construction and the empirical audit rules (challenge \hyperref[chal:2]{2}).
Section \ref{tool} presents the tool implementation and the RFC for community standardization (challenge \hyperref[chal:3]{3}).
Finally, Section \ref{sec:std-lib} evaluates the effectiveness of our methodology through an empirical audit of the Rust standard library.

\section{Unsafety Genesis and Annotation}\label{sec:annotation}
To formalize safety requirements, we analyze the official undefined behavior (UB) documentation and the safety comments in the Rust standard library. Based on the UB rules, we derive 21 safety tags across five categories to represent these requirements.

\subsection{Characterizing Undefined Behaviors in Rust}\label{sec:ub}
According to the official Rust specification in the Rust Nomicon~\cite{working-with-unsafe} and the Rust Reference~\cite{ub}, programmers should avoid undefined behavior. If a program executes an undefined operation, the compiler's correctness guarantees become invalid.
To establish a basis for our safety tag derivation, we classify undefined behaviors from these specifications into four categories, each encapsulating a core dimension of Rust’s safety model, as summarized in Table~\ref{tab:ub-taxonomy}.

\subsubsection{Memory and Pointer Safety.}
This category is the foundational layer of Rust's safety model, governing low-level memory accesses and pointer manipulations.
Violations in this domain include accessing dangling pointers or memory outside a single live allocation, performing misaligned memory accesses, exceeding valid bounds during pointer arithmetic or collection indexing, and permitting unsynchronized concurrent modifications that lead to data races in shared memory.

\subsubsection{Aliasing Rule Violations.}
Aliasing Rule Violations target Rust's unique ownership system, which the compiler leverages to enable strong optimizations and safety guarantees. This encompasses breaches of the exclusivity guarantee for mutable references (\&mut T) and the immutability invariant for shared references (\&T), except within the \texttt{UnsafeCell}. 

\subsubsection{Invalid Value Production}
Invalid Value Production addresses type-level validity invariants that must be maintained throughout program execution. This includes creating values with illegal bit patterns (such as invalid \texttt{bool}, \texttt{char}, or enum discriminants) and reading uninitialized memory for types with restricted value sets. 

\subsubsection{Runtime and External Interaction}
This category covers violations of the execution environment's contracts, including foreign function interface (FFI) conventions, compiler intrinsics, platform-specific features, and runtime assumptions.

\begin{table}[t]
\centering
\caption{Taxonomy of Undefined Behaviors in Rust.}
\label{tab:ub-taxonomy}
\rowcolors{2}{gray!10}{white}
\resizebox{\textwidth}{!}{
\begin{tabular}{ll}
\toprule
\rowcolor{gray!20}
\textbf{Category} & \textbf{Specific Behaviors} \\
\midrule
\cellcolor{blue!10}Memory/Pointer Safety  & \cellcolor{blue!5}Dangling and Illegal Access / \cellcolor{blue!5}Alignment Violation / \cellcolor{blue!5}Out-of-Bounds Access / \cellcolor{blue!5}Data Race \\
\cellcolor{green!10}Aliasing Rules & \cellcolor{green!5} Pointer Aliasing Violation / Immutable Mutation \\
\cellcolor{orange!10}Value Validity & \cellcolor{orange!5} Invalid Values / Uninitialized Memory \\
\cellcolor{red!10}Runtime/External Interaction & \cellcolor{red!5} FFI Violation / Platform Features / Runtime Violations / Inline Assembly \\
\bottomrule
\end{tabular}
}
\end{table}

\subsection{Derivation of Safety Tags}
\solution{Solution to Challenge \hyperref[chal:1]{1}} 
We introduce a tag-derivation workflow to extract safety tags from natural-language documentation. This process uses manual semantic analysis to translate informal comments into formal contracts based on Rust's memory model.

\subsubsection{Constraint Extraction and Decoupling.}
We reviewed the public unsafe API documentation in the standard library. To resolve nested or implied constraints, we traced cross-references to their base definitions. For example, the ambiguous phrase "pointer must be valid" resolves to \textit{dereferenceable} for non-zero-sized types. This property further breaks down into specific constraints like "must not be dangling" and "must lie within the allocation range".

\subsubsection{Taxonomy Mapping}
We categorized these constraints based on the UB types they prevent (Table \ref{tab:ub-taxonomy}). This resulted in a five-part taxonomy: \texttt{Layout}, \texttt{Pointer}, \texttt{Content}, \texttt{Alias}, and \texttt{Contextual}. We separated \texttt{Layout} from \texttt{Pointer} for finer granularity, and aligned \texttt{Content} and \texttt{Alias} with value validity and aliasing rules. Additionally, we introduced the \texttt{Contextual} category for state-dependent invariants, replacing the rarely used external category.

\subsubsection{Bottom-Up Synthesis}
We grouped recurring constraints into unified safety tags by a bottom-up analysis. For example, \texttt{core::num::unchecked\_add} requires that \texttt{self + rhs} does not overflow, while \texttt{core::slice::from\_raw\_parts} requires the slice size to not exceed \texttt{isize::MAX}. Since both enforce numerical bounds, we represent them with a single \textit{ValidNum} tag. 
All derived tags were cross-validated by two authors to ensure consistency.

\subsubsection{Parameterized Specification}
Finally, we add context-specific arguments to the tags to define the exact safety requirements.
To illustrate, the \texttt{ValidNum} tag can be parameterized as \text{ValidNum(self + rhs <= u8::MAX)} and \text{ValidNum(len * size\_of::<T>() <= isize::MAX)}, codifying the specific safety requirements in different contexts.

\subsection{Taxonomy of Safety Tags}

\begin{table*}[t]
\centering
\caption{Primitive safety properties.}
\label{tab:sp}
\resizebox{0.9\textwidth}{!}{%
\begin{tabular}{c|c|c|c}
\toprule
\rowcolor{gray!20}
\textbf{Category} & \textbf{Safety Tag} & \textbf{Expression} & \textbf{Role} \\ \toprule
\multirow{3}{*}{\begin{tabular}[c]{@{}c@{}} Layout \end{tabular}} &
 \texttt{Align} & $\rho \% \text{align}(\tau) = 0 \land \text{sizeof}(\tau) \% \text{align}(\tau) = 0$ & precondition \\ 
& \texttt{Size} & $\text{sizeof}(\tau) \in \mathcal{S}$ & precondition, option \\ 
& \texttt{!Padding} & $\text{padding}(\tau) = 0$ & precondition \\ \midrule
\multirow{4}{*}{\begin{tabular}[c]{@{}c@{}} Pointer \end{tabular}}
& \texttt{!Null} & $\rho \neq \mathbf{0}$ & precondition \\
& \texttt{Allocated} & $\forall i \in [0, \text{sizeof}(\tau) \times \lambda), \text{allocator}(\rho + i) = \alpha$ & precondition \\
& \texttt{InBound} & $\text{mem}[\rho, \rho + \text{sizeof}(\tau) \times \lambda) \subseteq \text{single\_allocated\_obj}$ & precondition \\
& \texttt{!Overlap} & $|\rho_d - \rho_s| > \text{sizeof}(\tau) \times \lambda$ & precondition \\
\midrule
\multirow{6}{*}{\begin{tabular}[c]{@{}c@{}} Content \end{tabular}} 
& \texttt{ValidNum} & $\epsilon \in \varrho \land \text{no\_overflow}(\epsilon)$ & precondition \\ 
& \texttt{ValidString} & $\text{mem}[\varrho_s] \subseteq \text{UTF-8}$ & precond, hazard \\
& \texttt{ValidCStr} & $\text{mem}[\rho + \lambda, \rho + \lambda + 1] = \text{null}$ & precondition \\
& \texttt{Init} & $\forall i \in [0, \lambda), \text{mem}[\rho + \text{sizeof}(\tau) \times i, \rho + \text{sizeof}(\tau) \times (i+1)) = \text{valid}(\tau)$ & precond, hazard \\
& \texttt{Unwrap} & $\text{unwrap}(\nu) = \tau$ & precondition \\
& \texttt{Typed} & $\text{typeOf}(*\rho) = \tau$ & precondition \\ \midrule
\multirow{3}{*}{\begin{tabular}[c]{@{}c@{}} Alias  \end{tabular}} & \texttt{Owning} & $\text{ownership}(*\rho) = \varnothing$ & precondition \\
& \texttt{Alias} & $\rho_1 = \rho_2$ & hazard \\
& \texttt{Alive} & $\text{lifetime}(*\rho) \geq \lambda_t$ & precondition \\ \midrule
\multirow{5}{*}{\begin{tabular}[c]{@{}c@{}} Contextual \end{tabular}} 
& \texttt{Pinned} & $\forall t \in [0, \lambda_t), \&_t(*\rho)_0 = \rho$ & hazard \\
& \texttt{!Volatile} & $\nexists \text{ thread } \theta, \theta.\text{write}(\rho, \rho + \text{sizeof}(\tau) \times \lambda)$ & precondition \\
& \texttt{Opened} & $\exists \text{ openfile}() \leadsto \phi \land \nexists \text{ closefile}(\phi)$ & precondition \\
& \texttt{Trait} & $\chi \in \text{traitimpl}(\tau)$ & option \\
& \texttt{!Reachable} & $\text{SAT}(\text{cond}()) = \text{false}$ & precondition \\
\bottomrule
\multicolumn{4}{l}{\footnotesize \textbf{Notation}: $\rho$: pointer, $\tau$: type, $\sigma$: size, $\lambda$: length, $\alpha$: allocator, $\epsilon$: expression, $\varrho$: value range, $\nu$: value, $\lambda_t$: lifetime, $\phi$: file descriptor, $\chi$: trait, $\mathbb{N}$: number,} \\
\multicolumn{4}{l}{\footnotesize \hspace{1.25cm}$\bot$: unknown, $\top$: any, $\theta$: thread, $\mathcal{S}$: a subset of $\mathbb{N}$} \\
\end{tabular}%
}
\end{table*}

\subsubsection{Layout}
The Rust reference book introduces three key attributes related to layout.

\textit{\uline{Alignment}}: 
While alignment violations are a form of pointer-related UB, we categorize alignment here as it is an integral part of a type's memory layout. An object of type \texttt{T} at address \texttt{p} is considered aligned if \texttt{p \% alignment(T) = 0}, and sizeof(\texttt{T}) must be a multiple of its alignment. These constraints are captured by the \texttt{Align} combined with arguments pointer \texttt{p} and type \texttt{T}. 

\textit{\uline{Size}}: 
The safety tag \texttt{Size} pertains to whether a type's size is known at compile time. It can be parameterized to specify further constraints. For instance, \texttt{Sized(T, any)} denotes a statically sized type, while \texttt{Sized(T, unknown)} indicates a dynamically sized type. A common requirement is that a type must not be a zero-sized type (ZST), as seen in \texttt{NonNull::offset\_from} method. This is expressed as \texttt{Size(T, !0)}, or \texttt{!ZST} as a variant or shorthand.

\textit{\uline{Padding}}: 
This property asserts that type \texttt{T} has no unused padding bytes and its safety tag can be presented as \texttt{!Padding}. This is crucial for operations like \texttt{raw\_eq()}, where padding contents could otherwise affect the comparison result.

\subsubsection{Pointer}
This category concerns pointer validity. 
The \texttt{!Null} property asserts that a pointer must not be null, a requirement for APIs like \texttt{NonNull::new\_unchecked}.
The \texttt{Allocated} property, required by functions such as \texttt{vec::from\_raw\_parts\_in}, ensures a pointer is not dangling and points to a valid memory region of a specified size and it must be allocated by the specific allocator \texttt{A}.
The property \texttt{InBound} ensures that the memory space is not only allocated but allocated for the same object's memory boundary. Finally, the property \texttt{!Overlap} ensures that the two memory spaces pointed to by \texttt{dst} and \texttt{src} do not overlap.

\subsubsection{Content}
This category ensures that memory contents adhere to type-specific invariants.
For instance, the generic-typed API, \texttt{Box::assume\_init(self) -> Box<[T], A>} requires the memory of the generic-typed value to be properly initialized before calling the method. For other concrete types, there are specific properties:

\textit{\uline{Valid Number}}: 
The \texttt{ValidNum} property ensures that the result of an arithmetic expression falls within the valid range of its numeric type (\textit{e.g.} \texttt{isize}, \texttt{i32}). This annotation also implies that no intermediate computations overflow. In its usage, additional elements can be added to specify the first parameter as an arithmetic expression, and the second as the valid value ranges.

\textit{\uline{Valid String}}: 
The \texttt{ValidString} property requires a memory region to contain a valid UTF-8 sequence, upholding Rust's string encoding invariants.

\textit{\uline{Valid CStr}}: 
The \texttt{ValidCStr} property mandates that a sequence of bytes is a valid C-style string, terminating with a null byte, as required by APIs like \texttt{CStr::from\_ptr}.

Furthermore, for enumerated types, the \texttt{Unwrap} safety tag specifies the variant required for a safe unchecked unwrap. For instance, \texttt{Option::unwrap\_unchecked(self)} can be annotated with \texttt{Unwrap(self, Some(T))}. Finally, \texttt{Typed} means that the memory pointed by the pointer can be used to create an object of type T. Rust has such safety requirements for several constructors named \texttt{from\_raw}, such as \texttt{Rc::from\_raw(ptr: *const T) -> Rc<T>} which requires the pointers be returned from an \texttt{Rc} object via the \texttt{Rc::into\_raw(this: Rc<T, A>) -> *const T} method.

\subsubsection{Alias}
This category relates to the context of ownership, borrowing, and lifetime rules enforced by the Rust compiler, which are fundamental to guaranteeing exclusive mutability. 
The \texttt{Owning} property is a precondition for APIs like \texttt{Box::from\_raw} that consume a raw pointer to create an owned value. It asserts that the pointer has no other owners to prevent double-free vulnerabilities~\cite{xu2021memory,li2021mirchecker}. 
In contrast, the \texttt{Alias} property addresses potential violations of mutable exclusivity, such as when converting a raw pointer to a reference like \texttt{pointer::as\_mut<'a>}. Unlike \texttt{Owning}, temporary violations of alias rules can be useful in certain scenarios (\textit{e.g.,} double-linked list). 
Thus, we classify the safety property of \texttt{Alias} as a \textit{hazardous} status, a post-requirement rather than a precondition (We will elaborate more on the classifications in Section~\ref{sec:spusage}). Finally, we use \texttt{Alive} to represent an object's lifetime must outlive a specified lifetime parameter.

\subsubsection{Contextual}
This category covers safety properties tied to specific execution contexts or semantic invariants. 
For instance, pinning is a characteristic that ensures that an object remains at the same memory address after being created and cannot be moved. Rust provides this feature via the \texttt{Pin} struct, and we annotate such a requirement for methods within this struct as \texttt{Pinned}. 
Another property related to memory access is volatile, which indicates that the memory may be subject to concurrent access and may change. In such cases, compilers should reload the value from memory rather than using the previous copy. Read or write operations via pointers (\textit{e.g.}, \texttt{ptr::read}) generally require the memory to be non-volatile. Otherwise, developers should use volatile methods. We represent this requirement with \texttt{!Volatile}. 
\texttt{Opened} is a precondition for I/O operations, requiring an underlying resource (\textit{e.g.}, a file descriptor for a \texttt{UdpSocket}) to be in an open state.
While distinct from direct memory-related UB, we categorize this as a safety property because it typifies a generalized class of \textit{system-level invariants} and is an intentional design of Rust (see RFC-3128~\cite{io_safety}).
Additionally, we use \texttt{Trait} to specify the condition that a type T implements a particular trait. Finally, \texttt{!Reachable} is a special requirement indicating that the current statement should not be reached during execution. The function \texttt{intrinsics::unreachable()} requires the safety property.

\subsubsection{Compound Safety Properties}
We define compound properties by combining primitive tags to concisely express recurring, complex safety requirements found in the Rust standard library documentation.

\texttt{\uline{Deref}}: This property guarantees the dereferencing of a pointer over a range from \texttt{p} to \texttt{p+Sizeof(T)*len} is safe. It consists of \texttt{Allocated} and \texttt{InBound}.
    
\texttt{\uline{ValidPtr}}: This property represents a pointer that is valid for memory access, as defined in libcore's \textit{ptr} module ~\cite{moduleptr}. The criteria distinguish zero-sized types (ZSTs) from others. For ZSTs, every pointer is valid. But for non-ZSTs, this requires the pointer to be dereferenceable, formulated as \texttt{ZST || (!ZST \&\& \texttt{Deref})}.
    
\texttt{\uline{Ptr2Ref}}: This property specifies the requirements for safely converting a raw pointer to a reference, and can be represented as \texttt{Align \&\& Deref \&\& Alias}.
    
\texttt{\uline{Layout}}: This property is used for memory allocation APIs (\textit{e.g.}, \texttt{Global::alloc(\&self, layout: Layout) -> *mut u8}). 
    It combines \texttt{ValidNum \&\& Allocated} to ensure a pointer is properly aligned and allocated according to a \texttt{Layout} struct.

\subsection{Role and Usage of Safety Properties} \label{sec:spusage}
Beyond their taxonomy and semantic arguments, safety properties are assigned specific contractual roles within an API's specification:

\textit{\uline{Precondition}}: A precondition is a safety property that constitutes a necessary condition for an API call to be sound. The caller is obligated to satisfy all preconditions collectively to establish the sufficient condition for avoiding UB. This is the most common role for safety properties.

\textit{\uline{Hazardous status (hazard)}}: 
A \textit{hazardous status} (or \textit{hazard}) denotes a persistent constraint imposed by an API call that the caller must uphold throughout subsequent operations. It functions as a post-requirement, distinct from a traditional \textit{postcondition} which only denotes a condition that must be true after the function call but does not require ongoing maintenance. For example, \texttt{!Init} can describe a hazardous state after calling \texttt{ptr::copy<T>} when the source memory is uninitialized.

\textit{\uline{Optional precondition (option)}}: 
An \textit{optional} precondition (\textit{option}) represents a sufficient but not necessary condition for safety.
Satisfying it ensures the safety of calling an unsafe API, but it is not mandatory. For example, \texttt{Pin::new\_unchecked(pointer: Ptr)} is guaranteed safe if the underlying type of the pointer implements the \textit{Unpin} trait (\texttt{Trait(T, Unpin)}), but this is not mandatory; if the type is not \textit{Unpin}, the caller must instead uphold the \texttt{Pinned} hazardous state.

Collectively, a full safety property comprises three elements: a tag name, its arguments, and its designated role.
This separation provides flexibility for different analysis scenarios. In lightweight code reviews, the tag name alone suffices to convey the core requirement. For more precise verification, arguments can be instantiated with concrete program variables using the notations from Table ~\ref{tab:sp}, such as \texttt{Init($\rho$,$\tau$,$\lambda$)}, which maintains readability by clarifying a constraint that the memory range, starting at \texttt{$\rho$} and spanning \texttt{sizeof($\tau$)*$\lambda$} bytes, must be properly initialized.
Furthermore, the role specifies whether this constraint must be verified before (\texttt{Precondition}), after (\texttt{Hazard}) or is optional (\texttt{Option}).
Thus, the initialization requirement for \texttt{slice::from\_raw\_parts(data: *const T, len: usize)} can be annotated as \texttt{\#[safety::requires(precond.Init(data,T,len))]}, where the prefix \texttt{precond.} is usually elided if it is the sole role of this property.

\subsection{Practical Applicability of the Tag-based Representation}
Programs have an infinite logic space, but any practical annotation system must be finite. Using a limited set of tags to cover complex program logic works in practice because unsafe code has two key characteristics: semantic recurrence and contractual abstraction.

First, semantic recurrence means that while code varies, the safety conditions required by low-level APIs frequently repeat standard patterns (such as pointer validity and alignment). This repetition allows a small set of tags to cover most cases. As demonstrated by our 96.1\% coverage of the standard library in Section~\ref{sec:std-lib}, a finite taxonomy is completely sufficient for real-world auditing.
Second, contractual abstraction separates high-level interface contracts from complex implementation details. Safety tags formalize high-level interface contracts.
For example, to handle diverse unchecked operations, the ValidNum tag can express completely different constraints simply by changing its parameters, \textit{e.g.}, \text{ValidNum(self + rhs <= u8::MAX)} and \text{ValidNum(len * size\_of::<T>() <= isize::MAX)}. This parameterization makes safety requirements generalizable across different APIs, keeping the tag taxonomy compact.
Consequently, highly specific, platform-dependent edge properties that defy such abstraction are intentionally excluded. Sacrificing exhaustive coverage of these long-tail properties in favor of a compact and generalizable tag system is a necessary trade-off for practical engineering.

\section{Auditing Methodology for Safety Documentation}\label{sec:method}
Building on the semantic foundation of safety tags, this section introduces a set of empirical rules to audit the consistency and completeness of API safety documentation.

\subsection{Motivating Example: Cross-Function Safety Dependencies}
This section uses \texttt{Box<T, A>} to illustrate how cross-function safety dependencies manifest in practice.

\begin{figure}
\footnotesize
\begin{subfigure}[b]{0.68\linewidth}
\begin{lstlisting}[language=Rust,style=boxed]
+ /// The pointer must point to the memory allocated by the global allocator.
pub unsafe fn from_raw(raw: *mut T) -> Box<T> {
  unsafe {Self::from_raw_in(raw, Global)}
}

+ /// The pointer must point to the memory allocated by alloc.
pub const unsafe fn from_raw_in(raw: *mut T, alloc: A) -> Box<T, A> { 
  Box(unsafe {Unique::new_unchecked(raw)}, alloc) 
}

pub fn new_in<A: Allocator>(x: T, alloc: A) -> Self {
    let mut boxed = Self::new_uninit_in(alloc);
    boxed.write(x);
    unsafe { boxed.assume_init() }
}

fn drop(&mut self) {
    let ptr = self.0;
    unsafe {
        let layout = Layout::for_value_raw(ptr.as_ptr());
        if layout.size() != 0 {
            self.1.deallocate(From::from(ptr.cast()), layout);
        }
    }
}
\end{lstlisting}
\end{subfigure}
\hfill
    \begin{subfigure}[b]{0.29\linewidth}
        \centering
        \includegraphics[width=\textwidth]{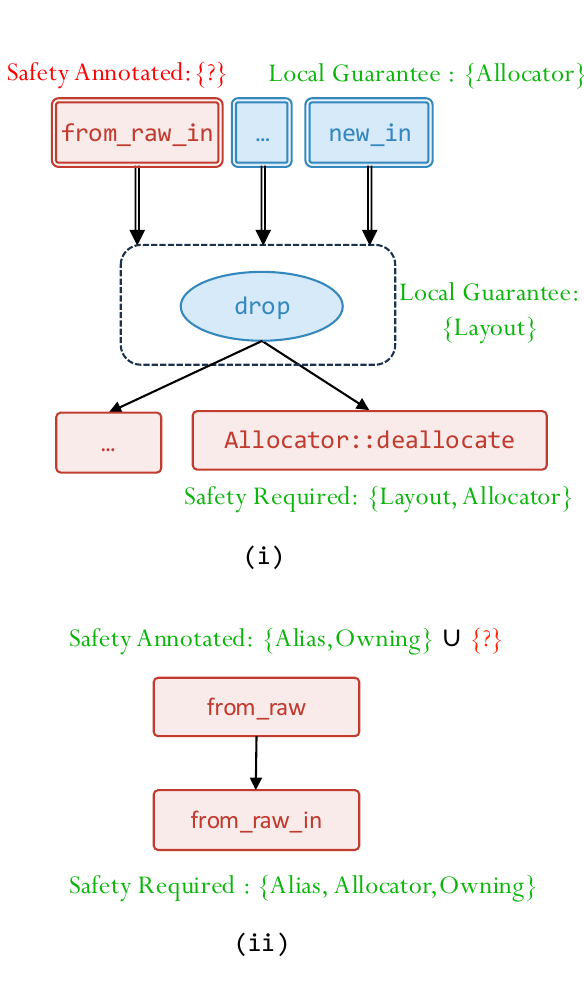}
    \end{subfigure}
  \caption{\textbf{Motivating examples} for safety documentation auditing and audit units in Box. (i) Method-level unit: the unsafe constructor \texttt{from\_raw\_in} fails to document the \texttt{Allocator} requirement propagated from \texttt{drop}. (ii) Function-level unit: \texttt{from\_raw} fails to delegate this requirement from \texttt{from\_raw\_in}.}
  \label{issue}
\end{figure}

As shown in Figure~\ref{issue}, \texttt{drop} is a safe function that calls the unsafe API \texttt{Allocator::deallocate}. The specification requires that the memory being deallocated must have been allocated by the specified allocator \texttt{A} (a requirement captured by the \texttt{Allocated} tag from Section ~\ref{sec:annotation}). Since \texttt{drop} performs no local validation, this requirement must be established at construction time.

\texttt{Box<T, A>} has multiple construction paths. 
For example, the safe constructor \texttt{new\_in} allocates memory via \texttt{A} internally, satisfying this prerequisite. However, the unsafe constructor \texttt{from\_raw\_in} accepts an externally provided raw pointer with no constraint on its allocator provenance, and its documentation contains no declaration of this requirement.
A caller can therefore pass a pointer allocated by a different allocator through \texttt{from\_raw\_in}.
When \texttt{drop} is subsequently triggered, the call to \texttt{deallocate} on that pointer produces undefined behavior.

This shows that safety requirements can span function boundaries.
From a program analysis perspective, this class of issues is conventionally attributed to the violation of struct invariants.
However, globally reasoning about struct invariants imposes substantial cost in large-scale audits.
We adopt a more operationally tractable perspective by reframing the audit as tracking how the safety tags are propagated, discharged, or omitted along the call chain or struct .
The following sections classify APIs based on their roles in the object lifecycle and define empirical rules to review the consistency and completeness of safety documentation across the entire state propagation chain.

\subsection{Audit Units}
Because unsafe code is encapsulated within various functional and structural boundaries, tracking safety requirements requires well-defined audit scopes. We introduce three core concepts to construct these boundaries.

\begin{definition}[Target API]
A target API is defined as any function or method that is explicitly marked \texttt{unsafe} or internally contains unsafe operations. It serves as the focal point for consistency verification, where documented safety tags are compared against actual implementation constraints.
\end{definition}

\begin{definition}[Constructor]
A Constructor is a function that instantiates a struct and does not consume an existing instance (\textit{i.e.}, no \texttt{self} parameter). Its return type logically instantiates the struct, encompassing both direct types (\textit{e.g.}, \texttt{Self}) and smart pointers (\textit{e.g.}, \texttt{Box<Self>}). Constructors act as the genesis of structural invariants.
\end{definition}

\begin{definition}[Audit Unit]\label{def:basic_units}
An Audit Unit is a minimal, self-contained scope designed to evaluate the documentation consistency of a specific Target API. It consists of a target API and its associated set of constructors.
\end{definition}

We classify Audit Units into two topological categories based on the presence of state coupling.

\textbf{Type I: Function-Level Unit.} 
A function-level unit is formed when the target API is a standalone function or a static method without a \texttt{self} reference.
The audit task is to check whether the internal code behaviors of the caller (\textit{e.g.}, \texttt{from\_raw} in Fig. ~\ref{issue}) satisfies the safety demands of the unsafe callees (\textit{e.g.}, \texttt{from\_raw\_in}), or whether it accurately propagates them to its own documented Safety Tags.

\textbf{Type II: Method-Level Unit.} 
A method-level unit is formed when the target API is an instance method of a struct (\textit{e.g.}, \texttt{drop}). This unit exhibits state coupling, meaning the safety may rely on the validity of the struct's internal fields.
Auditors should first examine whether the internal behaviors of the method completely handle the safety requirements of callee (\textit{e.g.}, \texttt{deallocate}), followed by checking whether all associated constructors (\textit{e.g.}, \texttt{new\_in} and \texttt{from\_raw\_in}) satisfy any missing safety requirements. 

\subsection{Audit Rules for Documentation Consistency}
\solution{Solution to Challenge \hyperref[chal:2]{2}}
We introduce three progressive audit rules to check documentation consistency in each audit unit: call-site check, constructor guarantee check, and state exposure check.

We audit safety documentation by checking that every safety obligation is either handled internally by code or explicitly documented for the user. We consider documentation correct if the safety requirements are stated directly at the target API (Rule \hyperref[rule1]{I}) or guaranteed by its constructors (Rule \hyperref[rule2]{II}). However, trusting constructor documentation requires the struct to protect its internal state from outside interference (Rule \hyperref[rule3]{III}). Ultimately, we report a documentation defect in two cases: a safety requirement is missing or incorrect from the docs, or the struct exposes its internal state, making its documented guarantees useless.

\vspace{1ex}
\phantomsection\label{rule1}\noindent\textbf{Rule I: Call-Site Check Rule}

This rule examines the target API directly. Any safety demands originating from its internal unsafe operations must be handled via internal code validation or explicitly declared in its public documentation. If a safety demand is neither programmatically discharged nor documented, it constitutes a documentation omission. For example, when auditing the function-level unit \texttt{from\_raw} in Fig.~\ref{issue}(ii), auditors identify that its internal callee (\texttt{from\_raw\_in}) inherently requires the \texttt{Allocator} property. Because \texttt{from\_raw} neither enforces this internally nor documents the \texttt{Allocator} tag for its callers, it violates this rule.



\vspace{1ex}
\phantomsection\label{rule2}\noindent\textbf{Rule II: Constructor Guarantee Check Rule}

When a target method relies on struct-level state, the audit evaluates its associated constructors. Every valid constructor must account for safety demands left unhandled or undocumented by the method. Constructors achieve this by either safely initializing the state internally or documenting the requirement for their own callers. If a method defers a safety obligation, and any constructor fails to initialize the required state while also omitting the safety documentation, the safety requirements are incomplete. For instance, in the method-level unit centered on \texttt{drop} in Fig.~\ref{issue}(i), the method calls \texttt{deallocate}, which demands the \texttt{Allocator} tag. Because \texttt{drop} does not handle this locally, the obligation transfers to the constructors. While the safe \texttt{new\_in} internally satisfies this requirement by allocating correctly, the unsafe \texttt{from\_raw\_in} neither enforces the allocator provenance nor documents the \texttt{Allocator} tag, violating this rule.



\vspace{1ex}
\phantomsection\label{rule3}\noindent\textbf{Rule III: State Exposure Check Rule}

This rule evaluates whether internal states governing an API's safety are exposed to external modifications (\textit{e.g.}, via public fields). When a struct exposes such critical states, the responsibility for maintaining safety invariants transfers to the caller. Thus, the exposed fields should explicitly document the safety constraints that callers are obligated to uphold. Without these documented restrictions, callers may inadvertently overwrite established invariants, rendering the struct's overall safety guarantees useless. For example, hypothetically assume \texttt{Box<T, A>} exposes its \texttt{alloc} field publicly. An external caller could replace the original allocator with a different one. When \texttt{drop} is subsequently invoked, it would attempt to deallocate the memory using the incorrect allocator, violating the \texttt{Allocator} property.




\section{Implementation}\label{tool}

\subsection{Operationalizing Safety Tags}

\solution{Solution to Challenge \hyperref[chal:3]{3}} 
To provide automated enforcement of safety tag consistency between caller annotations and callee requirements, we developed \textit{safety-tool}. This tool supports safety comments with \textit{attributes}: \texttt{\#[safety::requires]} and \texttt{\#[safety::checked]}.
As illustrated in Figure ~\ref{fig:tool},
\textit{safety-tool} connects developer annotations, tag definitions, and API contracts.

\subsubsection{Tag Definition and Configuration}
\begin{figure}[t]
    \centering
    \includegraphics[width=\linewidth]{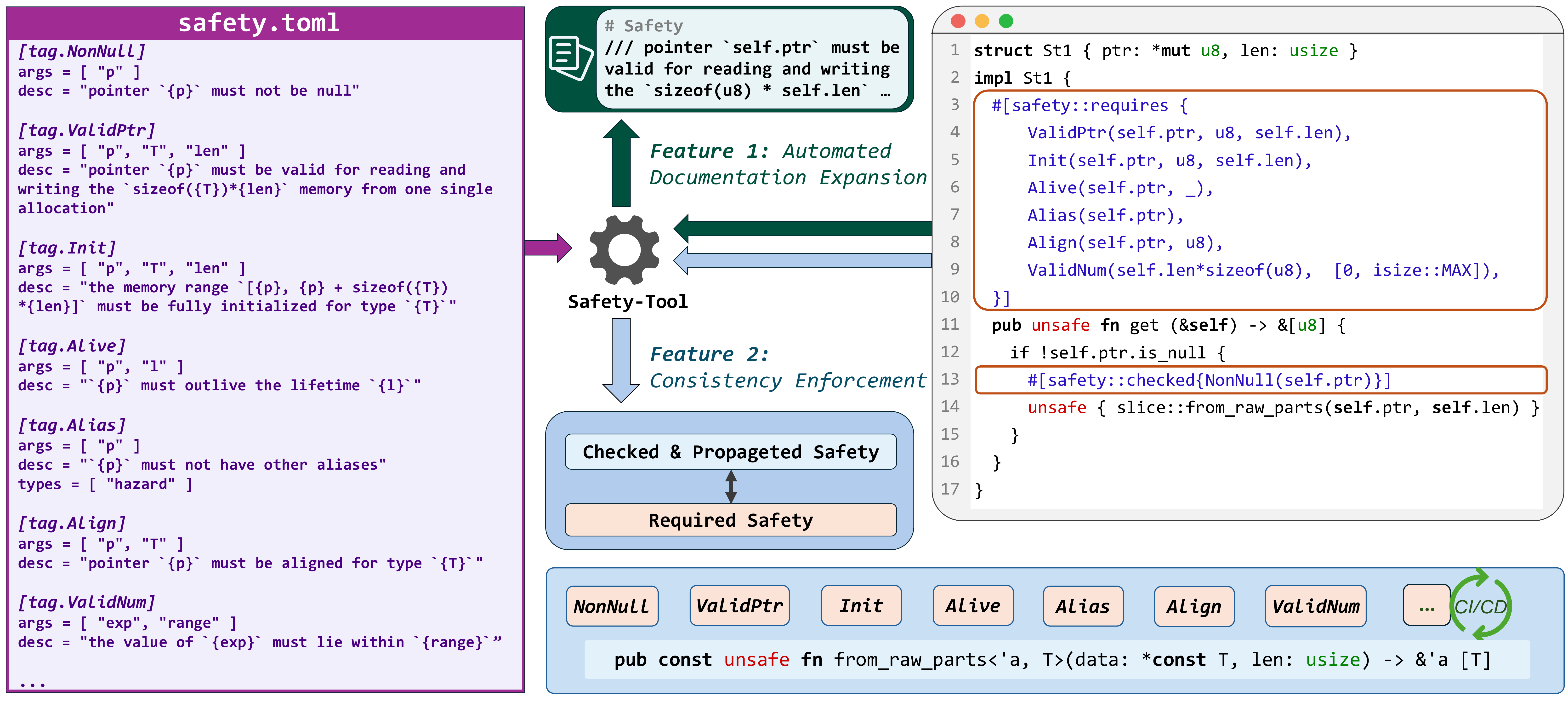}
    \caption{\textbf{The workflow and core features of \textit{safety-tool}}. The tool facilitates (1) \textbf{Automated Documentation Expansion}, which transforms structured safety tags into standard human-readable documentation; and (2) \textbf{Consistency Enforcement}, which statically verifies that the safety obligations of unsafe callees are strictly discharged or propagated by the caller.}
    \label{fig:tool}
\end{figure}

A safety tag is an attribute applied to both the definition and the call sites of unsafe functions. It corresponds to a single safety requirement consisting of a property identifier, arguments, roles, description, formal expression, and URL. Common properties are specified in a centralized TOML configuration file. The left side of Fig~\ref{fig:tool} shows the tag specification of \texttt{slice::from\_raw\_parts}. The identifier names the requirement, while arguments allow developers to plug in context-specific variables (\textit{e.g.}, \texttt{self.ptr}) to accurately express the safety constraint.

\subsubsection{Enforcement and Consistency}
Operating as a static linter, \textit{safety-tool} enforces consistency between a caller's annotations and the safety tags of its callees.
When calling a tagged unsafe function, the caller must provide corresponding \texttt{\#[safety::checked]} and \texttt{\#[safety::requires]} tags that include all defined property identifiers, optionally with justifications, to fully discharge or propagate the function's safety requirements (see the right side of Fig~\ref{fig:tool}). If any tags are missing, the tool will report an error and explain the unhandled requirements. This mechanism is particularly important for handling the evolution of external API safety contracts. For instance, if an updated dependency introduces a new safety requirement, the tool automatically identifies the discrepancy and issues an alert, allowing developers to catch latent safety regressions without manual re-auditing.

\subsubsection{Automated Documentation Expansion}
The tool can also expand documentation automatically by using a procedural macro to expand \texttt{\#[safety::requires]} into standard \texttt{\#[doc]} attributes. Instead of writing ad-hoc comments for an API, developers simply list the required tags (\textit{e.g.}, \texttt{NonNull}, \texttt{ValidPtr}, \texttt{Init}). The tool combines these tags with the TOML definitions to generate standardized, human-readable documentation directly in the source code. This ensures the documentation always stays in sync with the actual safety contracts.

\subsubsection{Implementation of safety-tool}
The tool is implemented as a Rust compiler driver, extracting attributes from unsafe functions and expressions. To enhance the developer experience, we also provide \textit{LSP plugins} for VSCode and Neovim, currently supporting autocompletion and hover documentation for safety tags.

\subsection{Integration with Rust Ecosystem}
To bring our idea to the entire Rust ecosystem, we have submitted a formal RFC~\cite{rust_rfc_3842} titled "Safety Tags" to the official Rust language repository. It pursues two concrete goals: (1) annotate every public unsafe API in the standard library with safety tags, and (2) extend Clippy to verify the tags and rust-analyzer to provide full language-server support for them. The RFC is presently under active deliberation, having accumulated over 50 commits and 170 comments to date. Several Rust projects have already shown initial interest in adopting safety tags (\textit{e.g.}, the Rust-for-Linux~\cite{rust_for_linux} and Bevy game~\cite{bevy_engine}).

\section{Experiments with The Rust Standard Library}\label{sec:std-lib}

To evaluate our methodology, we conducted a comprehensive audit of the Rust standard library,
encompassing the \texttt{core}~\cite{core}, \texttt{alloc}~\cite{alloc}, and \texttt{std}~\cite{std} crates. The target Rust version is nightly 1.84.0.
Two authors, with 3 and 6 years of Rust experience respectively, manually cross-reviewed the APIs.
To avoid redundancy, re-exported APIs were counted only once. 
In this section, we report the coverage of safety tags and the patterns of audit units, and detail the documentation inconsistencies identified during the audit.

\subsection{Safety Profile of Rust Standard Library}
Our audit of the Rust standard library identified 516 public unsafe APIs (excluding most intrinsic functions). In total, our safety tags can express the safety properties for 96.1\% of these APIs. 
The remaining 3.9\% pertain to system-level properties or functional correctness beyond the scope of our annotation language (\textit{e.g.}, \texttt{env::set\_var()}).
We only focus on a subset of intrinsic functions, excluding the majority whose safety (in terms of resolving to the correct linker symbols) is a platform requirement, not one that can be guaranteed by the caller. More details are provided in Section~\ref{sec:sp mission}. 
The \texttt{core}, \texttt{alloc}, and \texttt{std} crates contain 414, 64, and 38 public unsafe APIs, respectively. We achieved annotation coverage rates of 98.3\%, 82.8\%, and 84.2\%, assigning a total of 744, 105, and 44 safety tags to each crate, including both primitive and compound tags. The coverage of \texttt{alloc} and \texttt{std} is relatively low because they contain a portion of unsafe APIs involving function correctness and system properties.

\begin{figure}[t]
    \begin{minipage}[t]{0.9\textwidth}
        \centering
        \includegraphics[width=1\textwidth]{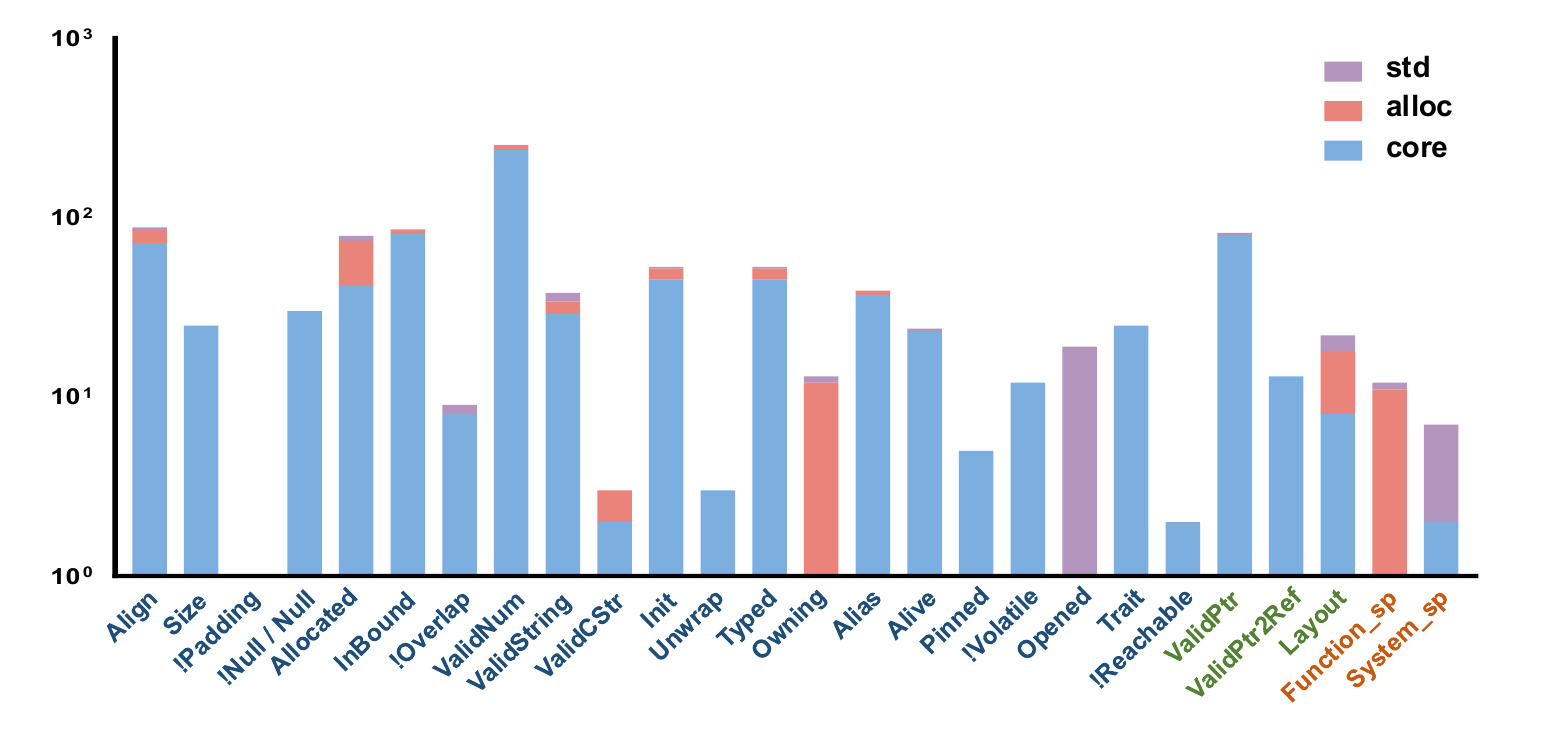}
        \caption{\textbf{Quantity distribution of safety property tags identified in the Rust standard library.}} 
        \label{fig:sp_data}
    \end{minipage}
\end{figure}

Figure~\ref{fig:sp_data} details the distribution of safety tags across the crates, plotted on a logarithmic scale due to the substantially higher tag count in the \texttt{core} crate. The \texttt{ValidNum} tag is predominant, reflecting its broad use in validating integer ranges (\textit{e.g.}, from \texttt{isize::MIN} to \texttt{isize::MAX}), indices, and offsets. 
The distribution also reveals crate-specific characteristics: The \texttt{alloc} crate relies more on \texttt{Owning} and function correctness (\textit{e.g.}, in collection types like \texttt{BTreeMap}), while the \texttt{std} crate is more constrained by \texttt{Opened} and other system properties related to the runtime environment (\textit{e.g.}, the unsafe APIs in the \texttt{env} module).

\begin{figure}[tbp]
\begin{minipage}[t]{0.88\textwidth}
  \centering
  \begin{subfigure}[b]{0.28\linewidth}
        \centering
        \includegraphics[width=\textwidth]{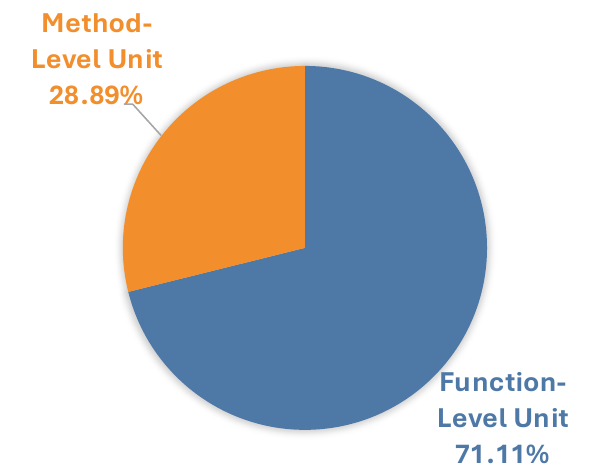}
        \caption{core}
        \label{fig:core_basic}
    \end{subfigure}
  \hfill 
  \begin{subfigure}[b]{0.28\linewidth}
        \centering
        \includegraphics[width=\textwidth]{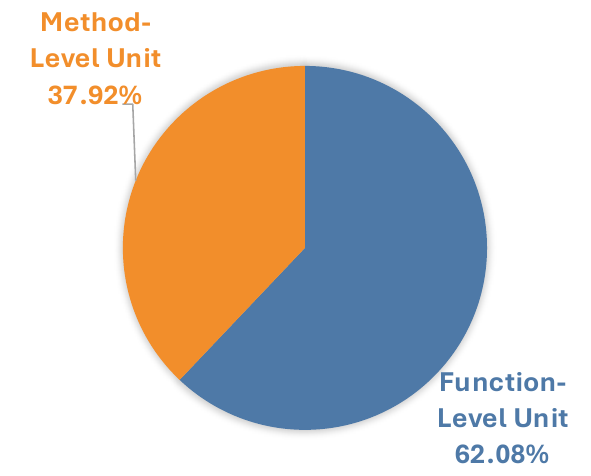}
        \caption{alloc}
        \label{fig:alloc_basic}
    \end{subfigure}
  \hfill
  \begin{subfigure}[b]{0.28\linewidth}
        \centering
        \includegraphics[width=\textwidth]{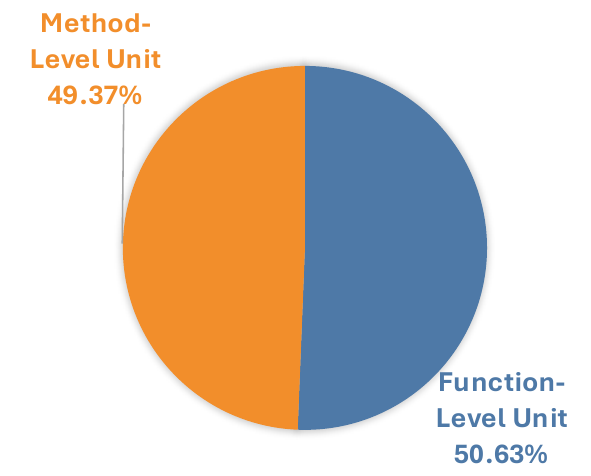}
        \caption{std}
        \label{fig:std_basic}
    \end{subfigure}
  \caption{\textbf{Distribution of unit patterns for each crate of the Rust standard library.}}
  \label{fig:basic_units}
  \end{minipage}
\end{figure}


We further analyze the structural characteristics of unsafe code usage by extracting audit units (Figure ~\ref{fig:basic_units}). The analysis yielded 1184 audit units for \texttt{core}, 269 for \texttt{alloc}, and 158 for \texttt{std}.
Both \texttt{alloc} and \texttt{std} contain a high proportion of method-level units. This reflects their roles in providing high-level abstractions and complex data structures (\textit{e.g.}, \texttt{Vec}, \texttt{Box}).
In sharp contrast, \texttt{core} is dominated by function-level units ($\approx$71.1\%). 
This ratio is primarily driven by the large number of primitive numeric APIs (\textit{e.g.}, for \texttt{u8}, \texttt{u16}) and intrinsic wrappers (\textit{e.g.}, \texttt{add\_unchecked}).
Crucially, this distribution serves as an indicator of audit complexity. Method-level units impose a higher audit burden, as they demand a review of multiple APIs.

\subsection{Tag-Based Safety Property Analysis}
\subsubsection{Inference Rules}
Applying the audit rules from Section~\ref{sec:method} requires significant manual effort.
To detect documentation defects at scale, our strategy is to filter APIs that show a high likelihood of documentation omissions. We achieve this by comparing the actual safety tags present in the documentation against an empirically derived set of expected tags.

To generate these expectations, we leverage the consistent naming conventions, type signatures, and interface patterns found in the Rust standard library. We implemented a script that codifies these conventions into a set of heuristic inference rules. For a given API, these rules deduce a predicted set of required safety obligations. A mismatch occurs when the inference rules demand a safety tag that is absent from the actual documentation. We flag APIs with such mismatches as highly suspicious candidates. We then manually investigate these candidates using our audit rules to confirm whether a true documentation defect exists.
This automated inference process does not strive for absolute formal completeness. Its primary objective is to act as a filter that identifies clear deviations from established ecosystem conventions.

To ensure our inference rules remain concise, we introduce the following notations: $f_u$ denotes an unsafe function/method; $P_f$, $R_f$, $N_f$, and $M_f$ denote the parameter set, return type, name, and module of function $f$, respectively. To replace legacy terminology, we define $\mathcal{T}_{req}(f)$ as the set of \textit{required} safety tags (obligations delegated to the caller).

\paragraph{\uline{Infer Safety Tags by API Signatures}}
By analyzing the functionality of Rust APIs in the standard library, we observe that certain API signatures strongly correlate with underlying safety properties. Furthermore, consistent naming conventions within crates can be leveraged to infer appropriate safety tags.

\begin{infrule} Bottom Annotation Rule. \label{infrule:bottom}
Any public unsafe API should specify a non-empty set of safety tags to declare its baseline safety requirements.

\footnotesize
    \begin{prooftree}
    \AxiomC{$f \text{ is pub } f_u$}
    \UnaryInfC{$\mathcal{T}_{req}(f) \neq \emptyset$}
    \end{prooftree}
\end{infrule}


\begin{infrule} Raw Pointer to Ownership. \label{infrule:raw2own}
APIs that construct owned objects from raw pointers should enforce alignment, allocation, bound, aliasing, and ownership constraints on the pointer.

\footnotesize
    \begin{prooftree}
    \AxiomC{$f \text{ is } f_u \land \text{raw\_ptr} \in P_f$}
    \AxiomC{$R_f \text{ is object} \lor \text{"from\_raw"} \in N_f$}
    \BinaryInfC{$\mathcal{T}_{req}(f) \supseteq \{\texttt{Align}, \texttt{Allocated}, \texttt{InBound}, \texttt{Alias}, \texttt{Owning} \}$}
    \end{prooftree}
\end{infrule}

\begin{infrule} Raw Pointer to Reference. \label{infrule:raw2ref}
APIs that convert raw pointers into references should require the pointer to satisfy alignment, allocation, bound, and aliasing constraints.

\footnotesize
    \begin{prooftree}
    \AxiomC{$f \text{ is } f_u \land \text{raw\_ptr} \in P_f$}
    \AxiomC{$R_f \text{ is reference} \lor \text{"as\_ref"} \in N_f$}
    \BinaryInfC{$\mathcal{T}_{req}(f) \supseteq \{\texttt{Align}, \texttt{Allocated}, \texttt{InBound}, \texttt{Alias} \}$}
    \end{prooftree}
 \end{infrule}

 \begin{infrule} Designated Allocator. \label{infrule:allocator}
 APIs accepting a specific allocator parameter should mandate that the underlying memory was originally allocated by that exact allocator.
 
 \footnotesize
    \begin{prooftree}
    \AxiomC{$f \text{ is } f_u \land \texttt{Allocator} \in P_f$}
    \AxiomC{$R_f \text{ is object} \lor \text{"\_in"} \in N_f$}
    \BinaryInfC{$\mathcal{T}_{req}(f) \supseteq \{\texttt{Allocated} \}$}
    \end{prooftree}
\end{infrule} 

 \begin{infrule} Unchecked Operations. \label{infrule:unchecked}
 APIs performing "unchecked" operations deduce module-specific safety tags, such as numerical validity, UTF-8 compliance, or boundary checks.
 \footnotesize
    \begin{prooftree}
    \AxiomC{$f \text{ is } f_u \land \text{"unchecked"} \in N_f$}
    \UnaryInfC{$
        \mathcal{T}_{req}(f) \supseteq 
        \begin{cases} 
            \{\texttt{ValidNum}\} & \text{if } M_f \in \text{integer modules} \\
            \{\texttt{ValidString}\} & \text{if } M_f = \texttt{str} \\
            \{\texttt{InBound}\} & \text{if } M_f = \texttt{slice}
        \end{cases}
    $}
    \end{prooftree}
\end{infrule} 

 \begin{infrule} Assume Init. \label{infrule:init}
APIs asserting memory initialization must explicitly require callers to ensure the target memory region is fully initialized.
 
 \footnotesize
    \begin{prooftree}
    \AxiomC{$f \text{ is } f_u \land \text{"assume\_init"} \in N_f$}
    \UnaryInfC{$\mathcal{T}_{req}(f) \supseteq \{\texttt{Init}\}$}
    \end{prooftree}
\end{infrule}

\paragraph{\uline{Safety Tag Analysis For Function Calls}}
According to the principle of encapsulation, all safety properties of unsafe callees must be either delegated or verified by the caller. Assuming a conservative approach where the caller verifies nothing, we establish the delegation rule:

\begin{infrule} Delegation Rule for Function Calls. \label{infrule:delegation}
Under a conservative audit assumption, the caller should delegate the required safety tags of all its unsafe callees to its documentation.
\footnotesize
    \begin{prooftree}
    \AxiomC{$\forall f_u \text{ invoked by } f$}
    \UnaryInfC{$\mathcal{T}_{req}(f) \supseteq \bigcup \mathcal{T}_{req}(f_u)$}
    \end{prooftree}
\end{infrule}

\subsubsection{Identified Safety Issues}
Overall, we have fixed the safety descriptions of 27 APIs, involving 61 safety tags. Table~\ref{tab:bugs} lists all these APIs along with the detailed information.

\begin{table}[t]
\centering
\setlength{\extrarowheight}{0pt}
\addtolength{\extrarowheight}{\aboverulesep}
\addtolength{\extrarowheight}{\belowrulesep}
\setlength{\aboverulesep}{0pt}
\setlength{\belowrulesep}{0pt}
\caption{\textbf{Fixed safety issues in the Rust standard library.} Allocator is a variant of \texttt{Allocated} with allocator parameter specified; \texttt{ZST} is a zero-sized variant of Sized.}
\label{tab:bugs}
\resizebox{\linewidth}{!}{%
\begin{tabular}{llllll} 
\toprule
\textbf{Type / Namespace} & \textbf{Function Signature} & \textbf{Fixed Tags} & \textbf{Issue Type} & \textbf{Aud. R.} & \textbf{Inf. R.} \\ 
\midrule
\textbf{Arc\textless{}\textcolor[rgb]{0.773,0,0}{T}\textgreater{}}  & \textbf{from\_raw(\textcolor[rgb]{0,0,0.51}{ptr}: \textcolor[rgb]{0,0.494,0}{*const }\textcolor[rgb]{0.773,0,0}{T}) -\textgreater{} Self}  & Allocator & Missing & Rule \hyperref[rule1]{I} & Rule \ref{infrule:delegation} \\
\rowcolor[rgb]{0.937,0.937,0.937} \textbf{Arc\textless{}\textcolor[rgb]{0.773,0,0}{T}\textgreater{}} & \textbf{increment\_strong\_count(\textcolor[rgb]{0,0,0.51}{ptr}: \textcolor[rgb]{0,0.494,0}{*const }\textcolor[rgb]{0.773,0,0}{T})} & Allocator \& ValidNum \& Aligned & Missing & Rule \hyperref[rule1]{I}  & Rule \ref{infrule:delegation} \\
\textbf{Arc\textless{}\textcolor[rgb]{0.773,0,0}{T}\textgreater{}} & \textbf{decrement\_strong\_count(\textcolor[rgb]{0,0,0.51}{ptr}: \textcolor[rgb]{0,0.494,0}{*const }\textcolor[rgb]{0.773,0,0}{T})} & Allocator \& ValidNum \& Aligned & Missing & Rule \hyperref[rule1]{I}  & Rule \ref{infrule:delegation} \\
\rowcolor[rgb]{0.937,0.937,0.937} \textbf{Arc\textless{}\textcolor[rgb]{0.773,0,0}{T}, \textcolor[rgb]{0.773,0,0}{A}\textgreater{}} & \textbf{increment\_strong\_count\_in(\textcolor[rgb]{0,0,0.51}{ptr}: \textcolor[rgb]{0,0.494,0}{*const }\textcolor[rgb]{0.773,0,0}{T}, \textcolor[rgb]{0,0,0.51}{alloc}: \textcolor[rgb]{0.773,0,0}{A})}  & ValidNum \& Aligned  & Missing & Rule \hyperref[rule1]{I}  & Rule \ref{infrule:delegation} \\
\textbf{Arc\textless{}\textcolor[rgb]{0.773,0,0}{T},\textcolor[rgb]{0.773,0,0}{ A}\textgreater{}} & \textbf{decrement\_strong\_count\_in(\textcolor[rgb]{0,0,0.51}{ptr}: \textcolor[rgb]{0,0.494,0}{*const }\textcolor[rgb]{0.773,0,0}{T}, \textcolor[rgb]{0,0,0.51}{alloc}: \textcolor[rgb]{0.773,0,0}{A})} & ValidNum \& Aligned   & Missing & Rule \hyperref[rule1]{I} & Rule \ref{infrule:delegation} \\
\rowcolor[rgb]{0.937,0.937,0.937} \textbf{Box\textless{}\textcolor[rgb]{0.773,0,0}{T}\textgreater{}}   & \textbf{from\_raw(\textcolor[rgb]{0,0,0.51}{raw}: \textcolor[rgb]{0,0.494,0}{*mut} \textcolor[rgb]{0.773,0,0}{T}) -\textgreater{} Self} & \begin{tabular}[c]{@{}l@{}}Allocator \\ Alias \end{tabular}  & \begin{tabular}[c]{@{}l@{}}Missing \\ Missing \end{tabular} & Rule \hyperref[rule1]{I}  & \begin{tabular}[c]{@{}l@{}} Rule \ref{infrule:delegation} \\ Rule \ref{infrule:raw2own}\end{tabular}  \\
\textbf{Box\textless{}\textcolor[rgb]{0.773,0,0}{T}\textgreater{}} & \textbf{from\_non\_null(\textcolor[rgb]{0,0,0.51}{ptr}: NonNull) -\textgreater{} Self} & \begin{tabular}[c]{@{}l@{}}Allocator \\ Alias \end{tabular} & \begin{tabular}[c]{@{}l@{}}Missing \\ Missing \end{tabular} & Rule \hyperref[rule1]{I}  & \begin{tabular}[c]{@{}l@{}} Rule \ref{infrule:delegation} \\ Rule \ref{infrule:raw2own} \end{tabular}  \\
\rowcolor[rgb]{0.937,0.937,0.937} \textbf{Box\textless{}\textcolor[rgb]{0.773,0,0}{T}, \textcolor[rgb]{0.773,0,0}{A}\textgreater{}}         & \textbf{from\_raw\_in(\textcolor[rgb]{0,0,0.51}{raw}: \textcolor[rgb]{0,0.494,0}{*mut} \textcolor[rgb]{0.773,0,0}{T}, \textcolor[rgb]{0,0,0.51}{alloc}: \textcolor[rgb]{0.773,0,0}{A}) -\textbf{\textgreater{}}~Self} & \begin{tabular}[c]{@{}l@{}}Allocator \\ Alias \end{tabular} & \begin{tabular}[c]{@{}l@{}}Missing \\ Missing \end{tabular} &  Rule \hyperref[rule1]{I} \& \hyperref[rule2]{II}  & \begin{tabular}[c]{@{}l@{}} Rule \ref{infrule:allocator} \\ Rule \ref{infrule:raw2own} \end{tabular}\\
\textbf{Box\textless{}\textcolor[rgb]{0.773,0,0}{T}, \textcolor[rgb]{0.773,0,0}{A}\textgreater{}}  & \textbf{from\_non\_null\_in(\textcolor[rgb]{0,0,0.51}{raw}: NonNull, \textcolor[rgb]{0,0,0.51}{alloc}: \textcolor[rgb]{0.773,0,0}{A}) -\textbf{\textgreater{}}~Self} & \begin{tabular}[c]{@{}l@{}}Allocator \\ Alias \end{tabular} & \begin{tabular}[c]{@{}l@{}}Missing \\ Missing \end{tabular} &  Rule \hyperref[rule1]{I} \& \hyperref[rule2]{II}  & \begin{tabular}[c]{@{}l@{}}Rule \ref{infrule:allocator} \\ Rule \ref{infrule:raw2own}\end{tabular}  \\
\rowcolor[rgb]{0.937,0.937,0.937} \textbf{Weak\textless{}\textcolor[rgb]{0.773,0,0}{T}\textgreater{}}  & \textbf{from\_raw(\textcolor[rgb]{0,0,0.51}{ptr}: \textcolor[rgb]{0,0.494,0}{*const} \textcolor[rgb]{0.773,0,0}{T}) -\textbf{\textbf{\textbf{\textgreater{}}}}~Self} & Allocator & Missing & Rule \hyperref[rule1]{I} & Rule \ref{infrule:delegation} \\
\textbf{CString} & \textbf{from\_raw(\textcolor[rgb]{0,0,0.51}{ptr}: \textcolor[rgb]{0,0.494,0}{*mut} c\_char) -\textgreater{} CString}  & Alias \& Owning \&  Allocated & Missing &  Rule \hyperref[rule1]{I} \& \hyperref[rule2]{II} & Rule \ref{infrule:raw2own} \\
\rowcolor[rgb]{0.937,0.937,0.937} \textbf{std::str}  & \textbf{from\_boxed\_utf8\_unchecked(\textcolor[rgb]{0,0,0.51}{v}: Box) -\textgreater{} Box}  & ValidString & Missing & Rule \hyperref[rule1]{I} \& \hyperref[rule2]{II} & Rule \ref{infrule:unchecked} \\
\textbf{Rc\textless{}\textcolor[rgb]{0.773,0,0}{T}\textgreater{}} & \textbf{increment\_strong\_count(\textcolor[rgb]{0,0,0.51}{ptr}: \textcolor[rgb]{0,0.494,0}{*const }\textcolor[rgb]{0.773,0,0}{ T})}  &   ValidNum \& Aligned & Missing & Rule \hyperref[rule1]{I}  & Rule \ref{infrule:delegation} \\
\rowcolor[rgb]{0.937,0.937,0.937} \textbf{Rc\textless{}\textcolor[rgb]{0.773,0,0}{T}\textgreater{}}  & \textbf{decrement\_strong\_count(\textcolor[rgb]{0,0,0.51}{ptr}: \textcolor[rgb]{0,0.494,0}{*const} \textcolor[rgb]{0.773,0,0}{T})} &   ValidNum \& Aligned & Missing & Rule \hyperref[rule1]{I} & Rule \ref{infrule:delegation} \\
\textbf{Rc\textless{}\textcolor[rgb]{0.773,0,0}{T}, \textcolor[rgb]{0.773,0,0}{A}\textgreater{}}  & \textbf{increment\_strong\_count\_in(\textcolor[rgb]{0,0,0.51}{ptr}: \textcolor[rgb]{0,0.494,0}{*const} \textcolor[rgb]{0.773,0,0}{T}, \textcolor[rgb]{0,0,0.51}{alloc}: \textcolor[rgb]{0.773,0,0}{A})} &  ValidNum \& Aligned & Missing & Rule \hyperref[rule1]{I} & Rule \ref{infrule:delegation} \\
\rowcolor[rgb]{0.937,0.937,0.937} \textbf{\textbf{Rc\textless{}\textcolor[rgb]{0.773,0,0}{T},~\textcolor[rgb]{0.773,0,0}{A}\textgreater{}}} & \textbf{decrement\_strong\_count\_in(\textcolor[rgb]{0,0,0.51}{ptr}: \textcolor[rgb]{0,0.494,0}{*const} \textcolor[rgb]{0.773,0,0}{T}, \textcolor[rgb]{0,0,0.51}{alloc}: \textcolor[rgb]{0.773,0,0}{A})} &   ValidNum \& Aligned  & Missing & Rule \hyperref[rule1]{I} & Rule \ref{infrule:delegation} \\
\textbf{std::ptr} & \textbf{read\_unaligned(\textcolor[rgb]{0,0,0.51}{src}: \textcolor[rgb]{0,0.494,0}{*const}\textcolor[rgb]{0.773,0,0}{ T}) -\textgreater{} \textcolor[rgb]{0.773,0,0}{T}}  & !Null || ZST & False & Rule \hyperref[rule1]{I}  & Rule \ref{infrule:delegation} \\
\rowcolor[rgb]{0.937,0.937,0.937} \textbf{std::ptr} & \textbf{write\_unaligned(\textcolor[rgb]{0,0,0.51}{src}: \textcolor[rgb]{0,0.494,0}{*const} \textcolor[rgb]{0.773,0,0}{ T}) -\textgreater{} \textcolor[rgb]{0.773,0,0}{T}} & !Null || ZST & False & Rule \hyperref[rule1]{I} & Rule \ref{infrule:delegation} \\
\textbf{intrinsics} & \textbf{typed\_swap\_nonoverlapping<\textcolor[rgb]{0.773,0,0}{T}>(\textcolor[rgb]{0,0,0.51}{x}: \textcolor[rgb]{0,0.494,0}{*mut} \textcolor[rgb]{0.773,0,0}{T}, \textcolor[rgb]{0,0,0.51}{y}: \textcolor[rgb]{0,0.494,0}{*mut} \textcolor[rgb]{0.773,0,0}{ T}) -\textgreater{} \textcolor[rgb]{0.773,0,0}{T}} & Typed \& ValidPtr \& Aligned & Missing & Rule \hyperref[rule1]{I} & Rule \ref{infrule:bottom} \\
\rowcolor[rgb]{0.937,0.937,0.937} \textbf{intrinsics} & \begin{tabular}[c]{@{}l@{}} \textbf{volatile\_copy\_nonoverlapping\_memory<\textcolor[rgb]{0.773,0,0}{T}>} \\ \textbf{  (\textcolor[rgb]{0,0,0.51}{dst}: \textcolor[rgb]{0,0.494,0}{*mut} \textcolor[rgb]{0.773,0,0}{T}, \textcolor[rgb]{0,0,0.51}{src}: \textcolor[rgb]{0,0.494,0}{*const} \textcolor[rgb]{0.773,0,0}{T}, \textcolor[rgb]{0,0,0.51}{count}: usize) } \end{tabular} & \begin{tabular}[c]{@{}l@{}} !Volatile \& ValidPtr \& Align \\ !Overlap \& Alias \& Trait \end{tabular} & Missing & Rule \hyperref[rule1]{I}  & Rule \ref{infrule:bottom} \\
\textbf{intrinsics} & \textbf{volatile\_set\_memory<\textcolor[rgb]{0.773,0,0}{T}>(\textcolor[rgb]{0,0,0.51}{x}: \textcolor[rgb]{0,0.494,0}{*mut} \textcolor[rgb]{0.773,0,0}{T}, \textcolor[rgb]{0,0,0.51}{y}: \textcolor[rgb]{0,0.494,0}{*mut} \textcolor[rgb]{0.773,0,0}{T}) } & \begin{tabular}[c]{@{}l@{}} !Volatile \& ValidPtr \\ Typed \& Align \end{tabular} & Missing & Rule \hyperref[rule1]{I}  & Rule \ref{infrule:bottom} \\
\rowcolor[rgb]{0.937,0.937,0.937} \textbf{intrinsics} & \textbf{va\_copy<\textcolor[rgb]{0.773,0,0}{'f}>} \textbf{( \textcolor[rgb]{0,0,0.51}{dest}: \textcolor[rgb]{0,0.494,0}{*mut} VaListImpl<\textcolor[rgb]{0.773,0,0}{'f}>, \textcolor[rgb]{0,0,0.51}{src}: \textcolor[rgb]{0,0.494,0}{\&}VaListImpl<\textcolor[rgb]{0.773,0,0}{'f}>)}   & !Null \& Allocated \& Alias & Missing & Rule \hyperref[rule1]{I} & Rule \ref{infrule:bottom} \\
\textbf{intrinsics} & \textbf{va\_arg<\textcolor[rgb]{0.773,0,0}{T}>(\textcolor[rgb]{0,0,0.51}{ap}: \textcolor[rgb]{0,0.494,0}{\&mut} VaListImpl<\textcolor[rgb]{0.773,0,0}{'\_}>) } & InBound \& Typed \& Init & Missing & Rule \hyperref[rule1]{I} \& \hyperref[rule2]{II} & Rule \ref{infrule:bottom} \\
\rowcolor[rgb]{0.937,0.937,0.937} \textbf{intrinsics} & \textbf{va\_end<\textcolor[rgb]{0.773,0,0}{T}>(\textcolor[rgb]{0,0,0.51}{ap}: \textcolor[rgb]{0,0.494,0}{\&mut} VaListImpl<\textcolor[rgb]{0.773,0,0}{'\_}>) } & Allocated & Missing & Rule \hyperref[rule1]{I} \& \hyperref[rule2]{II} & Rule \ref{infrule:bottom} \\
\textbf{VaListImpl}  & \textbf{arg(\textcolor[rgb]{0,0.494,0}{\&mut} \textcolor[rgb]{0,0,0.51}{self}) -\textgreater{} \textcolor[rgb]{0.773,0,0}{T}}  & InBound \& Typed \& Init & Missing & Rule \hyperref[rule1]{I}  & Rule \ref{infrule:delegation} \\
\rowcolor[rgb]{0.937,0.937,0.937} \textbf{str}  & \textbf{from\_utf8\_unchecked\_mut(\textcolor[rgb]{0,0,0.51}{v}: \textcolor[rgb]{0,0.494,0}{\&mut} [u8]) -\textgreater{} \textcolor[rgb]{0,0.494,0}{\&mut} str}  & ValidString & Missing & Rule \hyperref[rule1]{I} \& \hyperref[rule2]{II} & Rule \ref{infrule:bottom} \\
\textbf{std::str}  & \textbf{from\_utf8\_unchecked\_mut(\textcolor[rgb]{0,0,0.51}{v}: \textcolor[rgb]{0,0.494,0}{\&mut} [u8]) -\textgreater{} \textcolor[rgb]{0,0.494,0}{\&mut} str}  & ValidString & Missing & Rule \hyperref[rule1]{I} \& \hyperref[rule2]{II}  & Rule \ref{infrule:bottom} \\
\bottomrule
\end{tabular}
}
\end{table}

\paragraph{\uline{Issues Detected via Inference Process}}~\label{sec:sp mission}
We have corrected root safety annotations for 14 APIs with 30 safety tags by inference rules. 
These documentation omissions were identified by comparing the APIs' existing documentation against the expected baselines generated by our inference rules, and subsequently applying the Audit Rule \hyperref[rule1]{I}.
Three of these were derived via value validity requirements. 
Two of them are named \texttt{str::from\_utf8\_unchecked\_mut}, one a static function and the other a method. 
The last one, \texttt{str::from\_boxed\_utf8\_unchecked}, converts a slice into the \texttt{String} type, which lacks a safety description that specifies that the input byte slice must contain valid UTF-8 data.
Five of these were detected based on Rust's ownership model, mandating an \texttt{Alias} tag when a raw pointer is converted into an owned object or a reference. For instance, the API \texttt{CString::from\_raw} constructs a CString from a raw pointer but previously omitted the aliasing constraint. This annotation is necessary because using the raw pointer while the CString is alive violates mutable exclusivity.
The remaining ones were from six unsafe APIs in the \texttt{core::intrinsics} module and were resolved using memory and pointer safety principles, such as \texttt{typed\_swap\_nonoverlapping}.
Beyond these, 171 public unsafe intrinsic APIs still lack safety descriptions. They are marked as unsafe to prevent UB associated with compiler platform features.
These intrinsic APIs are generally annotated with \texttt{\#[rustc\_intrinsic\_must\_be\_overridden]}, indicating that their implementations are not provided within the \texttt{core} crate so that they require platform-specific linkers. Although we have reported such issues, they cannot be easily addressed at this stage due to a lack of consensus on how to document their safety requirements or even whether such documentation is necessary, since these intrinsic APIs are not intended for general developer use.

\paragraph{\uline{Issues Detected via Call-Site and Constructor Checks.}}
Our application of audit rules uncovered a total of 29 safety property issues spanning 15 APIs.
Two of these were identified and fixed via the Audit Rule \hyperref[rule2]{II} (struct soundness and constructor consistency rules), which stipulate that safety properties not fully encapsulated by a method must be enforced by all of its constructors.
These issues, detected in \texttt{Box::from\_raw\_in} and \texttt{Box::from\_non\_null\_in}, originated from the audit unit of \texttt{Box<T,A>::drop} (see Figure~\ref{issue}). The \texttt{drop} method calls \texttt{A.deallocate}, which requires the memory is originally allocated by allocator A (i.e., the \textit{Allocator} safety property). However, the internal logic of \texttt{drop} does not enforce this property. To avoid the endogenous unsoundness, relevant constraints must be imposed in endogenous regions. As the \texttt{drop} is declared safe, this safety property must be propagated to the constructors. While the constructors such as \texttt{new\_in} establish the \textit{Allocator} safety property in their internal code logic by allocating via \texttt{A}, \texttt{from\_raw\_in} internally only calls \texttt{Unique::new\_unchecked()} and a literal \texttt{alloc} assignment, without imposing sufficient constraints. Therefore, the \textit{Allocator} safety requirement must be explicitly annotated.
Our audit units also revealed 23 instances of insufficient safety property delegation across 11 APIs.
According to Audit Rule \hyperref[rule1]{I}, the safety properties of an unsafe callee should be delegated by its caller if they are not verified in function calls. Most of the detected issues appear in four widely used structs: \texttt{Box}, \texttt{Rc}, \texttt{Arc} and \texttt{Weak}. Figure~\ref{issue} also demonstrates a sample issue in \texttt{Box::from\_raw}, which lacks the required \texttt{Allocator} safety property. 
The function internally passes the \texttt{Global} allocator to \texttt{Box::from\_raw\_in}, implying a need for allocator consistency.
The remaining one is \texttt{ffi::va\_list::arg}, which must uphold the safety properties of \texttt{intrinsics::va\_arg}.
In addition to missing safety properties, we also identified two false annotations related to pointer validity.
Specifically, two unsafe APIs that read or write memory based on raw pointer input unnecessarily require the \texttt{!Null} tag, even for pointers to zero-sized types. This requirement contradicts the latest specification in the documentation of pointer validity.
To fix this issue, the pointer should either be non-null or point to a value of a zero-sized type.

\paragraph{\uline{Issues Detected via State Exposure Checks}}
We found one issue by Audit Rule \hyperref[rule3]{III}. Overall, there are 320 data structures, including 147 public ones and 173 private ones. Among them, 42 public data structures have unsafe constructors, and 15 of these allow literal constructors. By applying Audit Rule \hyperref[rule3]{III}, we identified that the struct \texttt{Pin<T>} exhibits a soundness issue related to allowing a literal constructor. In particular, the data structure has an unsafe constructor, \texttt{new\_unchecked(pointer: Ptr) -> Pin<Ptr>}, which accepts a pointer to some data of a type that may or may not implement the \texttt{Unpin} trait (a marker trait indicating that the data can be safely moved). The constructor requires callers to uphold the associated safety property, \textit{i.e.,} either the pointed object implementing \texttt{Unpin} or ensuring the pinned object is not moved afterward. However, the data structure contains only a single pointer field, which is public, thereby enabling the use of a literal constructor. As a result, it is possible to bypass the intended safety requirements through the literal constructor.

We have reported the issue to the official Rust repository, but it remains unresolved and is therefore omitted from the table~\ref{tab:bugs}. The literal constructor remains enabled because it is required by the macro constructor \texttt{pin!}. Currently, two countermeasures have already been employed to mitigate the misuse of the literal constructor. First, the pointer field is marked with \texttt{\#[doc(hidden)]} to hide it from doc rendering and discourage direct access, although it remains publicly accessible if developers deliberately access it. Second, developers must enable the \texttt{unsafe\_pin\_internals} feature to use the literal constructor directly. However, these restrictions can still be bypassed through the macro constructor \texttt{pin!}. The issue also highlights the urgent need to establish a well-documented policy to ensure API soundness. As the ecosystem continues to grow, relying on ad hoc safety handling techniques can increase the risk of subtle, hard-to-detect vulnerabilities.

\section{Related Work}
\subsection{Safety Mechanisms and Research Related to Unsafe Rust}
To ensure safety, the Rust ecosystem employs Miri~\cite{miri} for dynamic UB detection, though it is limited by incomplete path coverage. Complementing this, compiler flags like \texttt{sanitizer}, \texttt{ub-checks}, and \texttt{contract-checks} enable runtime monitoring. The Rust Foundation's \textit{verify-rust-std} project~\cite{verify-std} further instruments the standard library with formal verification tools to address 27 unsafe API challenges. Kani is the well-established verifier in the project~\cite{vanhattum2022verifying}, owning 1369 harnesses and thousands of autoharnesses. Verifiers such as  ESBMC~\cite{gadelha2018esbmc}, VeriFast~\cite{jacobs2011verifast}, and Flux~\cite{flux} have also been acknowledged for their reliability.
In academia, research spans empirical studies on unsafe code~\cite{evans2020rust, astrauskas2020programmers} and memory safety bugs~\cite{qin2020understanding, xu2021memory, li2024empirical}. Technical approaches in this domain include static analysis for bug detection~\cite{cui2023safedrop, bae2021rudra, li2021mirchecker} and formal verification for ensuring soundness~\cite{jung2017rustbelt, matsushita2022rusthornbelt, lehmann2023flux, gaher2024refinedrust}.

\subsection{Specification Generation and Safety Property Analysis}
Research in specification generation has transitioned from traditional rule-based methods to modern LLM-driven approaches. Early efforts, such as Toradocu~\cite{goffi2016automatic}, JDoctor~\cite{blasi2018translating}, and various NLP-based tools~\cite{zhou2017analyzing, ernst2001dynamically}, focused on extracting functional constraints from documentation to synthesize test oracles for Java and C/C++. More recently, new frameworks like SpecRover~\cite{ruan2024specrover} and LLMxCPG~\cite{lekssays2025llmxcpg} further integrate structured program analysis with LLM to enhance auditing dependability and specification precision.
Distinct from these functional-oriented works, this paper focuses on Rust’s safety soundness by establishing a systematic safety tag taxonomy, providing a formal foundation for future automated generation.

Regarding safety property enforcement, typestate analysis~\cite{strom2012typestate} serves as a foundational technique for verifying state-dependent properties, extending even to object protocols~\cite{bierhoff2007modular}. Besides, Xu \textit{et al.}~\cite{xu2000safety} pioneered its application to untrusted machine code, ensuring memory safety by statically tracking fine-grained states (\textit{e.g.}, initialized vs. uninitialized) and linear constraints. Alongside graph-based vulnerability discovery methods like the code property graph (CPG)~\cite{yamaguchi2014modeling, yamaguchi2015automatic}, we synthesize all these philosophies into our auditing methodology.

\section{Conclusion}
This work addresses the problem of auditing unsafe Rust code in a practical and effective way. 
To tackle the challenge of analyzing safety properties of unsafe code, we present a novel unsafety propagation approach, featuring hazardous states of calling unsafe APIs over traditional post-conditions.
Our experiment shows that 96.1\% of unsafe APIs in the standard library are covered by all identified tags, and safety documentation issues were fixed for 27 APIs.
We also introduce an auditor named \textit{safety-tool} to formulate safety properties as structured, machine-readable comments and enforce the corresponding checks. An RFC titled “Safety Tags” has been submitted to the Rust project repository for official support and community review.
Moreover, UPG analysis and tag-based approach can be extended further: parameterized safety tags provide a ready springboard for automated formal verification in the future. 


\bibliographystyle{ACM-Reference-Format}
\bibliography{ref.bib}

@inproceedings{hind2001pointer,
  title={Pointer analysis: haven't we solved this problem yet?},
  author={Hind, Michael},
  booktitle={Proceedings of the 2001 ACM SIGPLAN-SIGSOFT workshop on Program analysis for software tools and engineering},
  pages={54--61},
  year={2001},
  publisher = {Association for Computing Machinery},
  address = {New York, NY, USA}
}

@inproceedings{astrauskas2022prusti,
  title={The prusti project: Formal verification for Rust},
  author={Astrauskas, Vytautas and B{\'\i}l{\`y}, Aurel and Fiala, Jon{\'a}{\v{s}} and Grannan, Zachary and Matheja, Christoph and M{\"u}ller, Peter and Poli, Federico and Summers, Alexander J},
  booktitle={NASA Formal Methods Symposium},
  pages={88--108},
  year={2022},
  organization={Springer},
  publisher = {Springer-Verlag},
  address = {Berlin, Heidelberg}
}

@inproceedings{sharma2024rust,
  title={Rust for Embedded Systems: Current State and Open Problems},
  author={Sharma, Ayushi and Sharma, Shashank and Tanksalkar, Sai Ritvik and Torres-Arias, Santiago and Machiry, Aravind},
  booktitle={Proceedings of the 2024 on ACM SIGSAC Conference on Computer and Communications Security},
  pages={2296--2310},
  year={2024},
  publisher = {Association for Computing Machinery},
  address = {New York, NY, USA}
}

@inproceedings{li2024empirical,
  title={An empirical study of Rust-for-Linux: The success, dissatisfaction, and compromise},
  author={Li, Hongyu and Guo, Liwei and Yang, Yexuan and Wang, Shangguang and Xu, Mengwei},
  booktitle={2024 USENIX Annual Technical Conference},
  pages={425--443},
  year={2024},
  address = {Santa Clara, CA},
  publisher = {USENIX Association}
}

@inproceedings{peng2024framekernel,
  title={Framekernel: A safe and efficient kernel architecture via Rust-based intra-kernel privilege separation},
  author={Peng, Yuke and Tian, Hongliang and Xian, Jinyi and Zhou, Shuai and Yan, Shoumeng and Zhang, Yinqian},
  booktitle={Proceedings of the 15th ACM SIGOPS Asia-Pacific Workshop on Systems},
  pages={31--37},
  year={2024},
  publisher = {Association for Computing Machinery},
  address = {New York, NY, USA}
}

@article{qin2024understanding,
  title={Understanding and detecting real-world safety issues in Rust},
  author={Qin, Boqin and Chen, Yilun and Liu, Haopeng and Zhang, Hua and Wen, Qiaoyan and Song, Linhai and Zhang, Yiying},
  journal={IEEE Transactions on Software Engineering},
  year={2024},
  volume={50},
  number={6},
  pages={1306-1324},
  publisher={IEEE}
}

@inproceedings{cui2024unsafe,
  title={Is unsafe an Achilles' Heel? A comprehensive study of safety requirements in unsafe Rust programming},
  author={Cui, Mohan and Sun, Shuran and Xu, Hui and Zhou, Yangfan},
  booktitle={Proceedings of the IEEE/ACM 46th International Conference on Software Engineering},
  pages={1--13},
  year={2024},
  publisher = {Association for Computing Machinery},
  address = {New York, NY, USA}
}

@article{cui2023safedrop,
  title={SafeDrop: Detecting memory deallocation bugs of rust programs via static data-flow analysis},
  author={Cui, Mohan and Chen, Chengjun and Xu, Hui and Zhou, Yangfan},
  journal={ACM Transactions on Software Engineering and Methodology},
  volume={32},
  number={4},
  pages={1--21},
  year={2023},
  publisher={ACM New York, NY, USA}
}

@inproceedings{vanhattum2022verifying,
  title={Verifying dynamic trait objects in Rust},
  author={VanHattum, Alexa and Schwartz-Narbonne, Daniel and Chong, Nathan and Sampson, Adrian},
  booktitle={Proceedings of the 44th International Conference on Software Engineering: Software Engineering in Practice},
  pages={321--330},
  year={2022},
  publisher = {Association for Computing Machinery},
  address = {New York, NY, USA}
}

@inproceedings{matsushita2022rusthornbelt,
  title={RustHornBelt: a semantic foundation for functional verification of Rust programs with unsafe code},
  author={Matsushita, Yusuke and Denis, Xavier and Jourdan, Jacques-Henri and Dreyer, Derek},
  booktitle={Proceedings of the 43rd ACM SIGPLAN International Conference on Programming Language Design and Implementation},
  pages={841--856},
  year={2022},
  publisher = {Association for Computing Machinery},
  address = {New York, NY, USA}
}

@inproceedings{qin2020understanding,
  title={Understanding memory and thread safety practices and issues in real-world Rust programs},
  author={Qin, Boqin and Chen, Yilun and Yu, Zeming and Song, Linhai and Zhang, Yiying},
  booktitle={Proceedings of the 41st ACM SIGPLAN Conference on Programming Language Design and Implementation},
  pages={763--779},
  year={2020},
  publisher = {Association for Computing Machinery},
  address = {New York, NY, USA}
}

@article{xu2021memory,
  title={Memory-safety challenge considered solved? An in-depth study with all Rust CVEs},
  author={Xu, Hui and Chen, Zhuangbin and Sun, Mingshen and Zhou, Yangfan and Lyu, Michael R},
  journal={ACM Transactions on Software Engineering and Methodology (TOSEM)},
  volume={31},
  number={1},
  pages={1--25},
  year={2021},
  publisher={ACM New York, NY}
}

@inproceedings{bae2021rudra,
  title={Rudra: Finding memory safety bugs in Rust at the ecosystem scale},
  author={Bae, Yechan and Kim, Youngsuk and Askar, Ammar and Lim, Jungwon and Kim, Taesoo},
  booktitle={Proceedings of the ACM SIGOPS 28th Symposium on Operating Systems Principles},
  pages={84--99},
  year={2021},
  publisher = {Association for Computing Machinery},
  address = {New York, NY, USA}
}

@inproceedings{li2021mirchecker,
  title={MirChecker: detecting bugs in Rust programs via static analysis},
  author={Li, Zhuohua and Wang, Jincheng and Sun, Mingshen and Lui, John CS},
  booktitle={Proceedings of The 2021 ACM SIGSAC Conference on Computer and Communications Security},
  pages={2183--2196},
  year={2021},
  publisher = {Association for Computing Machinery},
  address = {New York, NY, USA}
}

@inproceedings{evans2020rust,
  title={Is \text{Rust} used safely by software developers?},
  author={Evans, Ana Nora and Campbell, Bradford and Soffa, Mary Lou},
  booktitle={Proceedings of the ACM/IEEE 42nd International Conference on Software Engineering},
  pages={246--257},
  year={2020},
  publisher = {Association for Computing Machinery},
  address = {New York, NY, USA}
}

@article{astrauskas2020programmers,
  title={How do programmers use unsafe \text{Rust}?},
  author={Astrauskas, Vytautas and Matheja, Christoph and Poli, Federico and M{\"u}ller, Peter and Summers, Alexander J},
  journal={Proceedings of the ACM on Programming Languages},
  volume={4},
  number={OOPSLA},
  pages={1--27},
  year={2020},
  publisher={ACM New York, NY, USA}
}

@inproceedings{gadelha2018esbmc,
  title={ESBMC 5.0: An industrial-strength C model checker},
  author={Gadelha, Mikhail R and Monteiro, Felipe R and Morse, Jeremy and Cordeiro, Lucas C and Fischer, Bernd and Nicole, Denis A},
  booktitle={Proceedings of the 33rd ACM/IEEE International Conference on Automated Software Engineering},
  pages={888--891},
  year={2018},
  publisher = {Association for Computing Machinery},
  address = {New York, NY, USA}
}

@inproceedings{jacobs2011verifast,
  title={VeriFast: A powerful, sound, predictable, fast verifier for C and Java},
  author={Jacobs, Bart and Smans, Jan and Philippaerts, Pieter and Vogels, Fr{\'e}d{\'e}ric and Penninckx, Willem and Piessens, Frank},
  booktitle={NASA Formal Methods Symposium},
  pages={41--55},
  year={2011},
  organization={Springer},
  publisher = {Springer-Verlag},
  address = {Berlin, Heidelberg}
}

@article{jung2017rustbelt,
  title={RustBelt: Securing the foundations of the Rust programming language},
  author={Jung, Ralf and Jourdan, Jacques-Henri and Krebbers, Robbert and Dreyer, Derek},
  journal={Proceedings of the ACM on Programming Languages},
  volume={2},
  number={POPL},
  pages={1--34},
  year={2017},
  publisher={ACM New York, NY, USA}
}

@article{lattuada2023verus,
  title={Verus: Verifying Rust programs using linear ghost types},
  author={Lattuada, Andrea and Hance, Travis and Cho, Chanhee and Brun, Matthias and Subasinghe, Isitha and Zhou, Yi and Howell, Jon and Parno, Bryan and Hawblitzel, Chris},
  journal={Proceedings of the ACM on Programming Languages},
  volume={7},
  number={OOPSLA1},
  pages={286--315},
  year={2023},
  publisher={ACM New York, NY, USA}
}

@inproceedings{lattuada2024verus,
  title={Verus: A practical foundation for systems verification},
  author={Lattuada, Andrea and Hance, Travis and Bosamiya, Jay and Brun, Matthias and Cho, Chanhee and LeBlanc, Hayley and Srinivasan, Pranav and Achermann, Reto and Chajed, Tej and Hawblitzel, Chris and others},
  booktitle={Proceedings of the ACM SIGOPS 30th Symposium on Operating Systems Principles},
  pages={438--454},
  year={2024},
  publisher = {Association for Computing Machinery},
  address = {New York, NY, USA}
}

@article{lehmann2023flux,
  title={Flux: Liquid types for rust},
  author={Lehmann, Nico and Geller, Adam T and Vazou, Niki and Jhala, Ranjit},
  journal={Proceedings of the ACM on Programming Languages},
  volume={7},
  number={PLDI},
  pages={1533--1557},
  year={2023},
  publisher={ACM New York, NY, USA}
}

@article{gaher2024refinedrust,
  title={RefinedRust: A type system for high-assurance verification of Rust programs},
  author={G{\"a}her, Lennard and Sammler, Michael and Jung, Ralf and Krebbers, Robbert and Dreyer, Derek},
  journal={Proceedings of the ACM on Programming Languages},
  volume={8},
  number={PLDI},
  pages={1115--1139},
  year={2024},
  publisher={ACM New York, NY, USA}
}

@misc{verify-std,
  title        = "The verify Rust standard library effort",
  author      = "Verify Rust Std Lib",
  url          = "https://model-checking.github.io/verify-rust-std/intro.html",
  year         = "2025",
  note         = "[Accessed: 2025-04-10]"
}

@misc{moduleptr,
  title = {Module ptr},
  author      = "Rust Documentation",
  url = "https://doc.rust-lang.org/nightly/std/ptr/index.html",
  note = "[Accessed: 2025-02-01]",
  year = 2025
}

@misc{miri,
  title        = "{Miri}",
  author      = "The Rust Team",
  url          = "https://github.com/rust-lang/miri",
  note         = "[Accessed: 2025-04-10]",
  year = 2025
}

@misc{working-with-unsafe,
  title        = "Working with unsafe",
  author      = "Rust Nomicon",
  url          = "https://doc.rust-lang.org/nomicon/working-with-unsafe.html",
  year         = "2025",
  note         = "[Accessed: 2025-04-28]"
}

@misc{unsafe-rust,
  title        = "Unsafe Rust",
  author      = "Rust Book",
  url          = "https://doc.rust-lang.org/book/ch20-01-unsafe-rust.html",
  year         = "2025",
  note         = "[Accessed: 2025-04-28]"
}

@misc{visibility,
  title        = "The Rust reference: visibility and privacy",
  author      = "Rust reference",
  url          = "https://doc.rust-lang.org/reference/visibility-and-privacy.html",
  year         = "2025",
  note         = "[Accessed: 2025-04-28]"
}

@misc{ub,
  title        = "The Rust reference: Behavior considered undefined",
  author      = "Rust reference",
  url          = "https://doc.rust-lang.org/reference/behavior-considered-undefined.html",
  year         = "2025",
  note         = "[Accessed: 2025-04-28]"
}

@misc{core,
  title        = "The Rust core library",
  author      = "Rust Standard Library",
  url          = "https://doc.rust-lang.org/nightly/core/index.html",
  year         = "2025",
  note         = "[Accessed: 2025-04-28]"
}

@misc{alloc,
  title        = "The Rust core allocation and collections library",
  author      = "Rust Standard Library",
  url          = "https://doc.rust-lang.org/nightly/alloc/index.html",
  year         = "2025",
  note         = "[Accessed: 2025-04-28]"
}

@misc{std,
  title        = "The Rust standard library",
  author      = "Rust Standard Library",
  url          = "https://doc.rust-lang.org/nightly/std/index.html",
  year         = "2025",
  note         = "[Accessed: 2025-04-28]"
}

@article{peng2025asterinas,
  author = {Peng, Yuke and Tian, Hongliang and Zhang, Junyang and Li, Ruihan and Chen, Chengjun and Jiang, Jianfeng and Xian, Jinyi and Wang, Xiaolin and Xu, Chenren and Zhou, Diyu and Luo, Yingwei and Yan, Shoumeng and Zhang, Yinqian},
  title = {ASTERINAS: a Linux ABI-compatible, rust-based framekernel OS with a small and sound TCB},
  year = {2025},
  numpages = {17},
  isbn = {978-1-939133-48-9},
  publisher = {USENIX Association},
  address = {USA}
}

@article{tianasterinas,
  title={Asterinas: A Rust-Based Framekernel to Reimagine Linux in the 2020s},
  author={Tian, Hongliang and Peng, Yuke and Luo, Yingwei and Yan, Shoumeng and Zhang, Yinqian}
}

@misc{ub_promise,
  title        = "Rust Glossary",
  author       = "Rust Group",
  url          = "https://rust-lang.github.io/unsafe-code-guidelines/glossary.html",
  year         = "2025",
  note         = "[Accessed: 2025-09-17]"
}

@misc{unsafe-fields,
  title        = "Unsafe fields",
  author       = "Rust RFC",
  url          = "https://github.com/rust-lang/rfcs/pull/3458",
  year         = "2025",  
  note         = "[Accessed: 2025-09-17]"
}

@article{flux,
  author = {Lehmann, Nico and Geller, Adam T. and Vazou, Niki and Jhala, Ranjit},
  title = {Flux: Liquid Types for Rust},
  year = {2023},
  issue_date = {June 2023},
  publisher = {Association for Computing Machinery},
  address = {New York, NY, USA},
  volume = {7},
  number = {PLDI},
  url = {https://doi.org/10.1145/3591283},
  journal = {Proc. ACM Program. Lang.},
  articleno = {169},
  numpages = {25},
}

@inproceedings{goffi2016automatic,
  title={Automatic generation of oracles for exceptional behaviors},
  author={Goffi, Alberto and Gorla, Alessandra and Ernst, Michael D and Pezz{\`e}, Mauro},
  booktitle={Proceedings of the 25th international symposium on software testing and analysis},
  pages={213--224},
  year={2016}
}

@article{ernst2001dynamically,
  title={Dynamically discovering likely program invariants to support program evolution},
  author={Ernst, Michael D and Cockrell, Jake and Griswold, William G and Notkin, David},
  journal={IEEE transactions on software engineering},
  volume={27},
  number={2},
  pages={99--123},
  year={2001},
  publisher={IEEE}
}

@inproceedings{zhou2017analyzing,
  title={Analyzing APIs documentation and code to detect directive defects},
  author={Zhou, Yu and Gu, Ruihang and Chen, Taolue and Huang, Zhiqiu and Panichella, Sebastiano and Gall, Harald},
  booktitle={2017 IEEE/ACM 39th International Conference on Software Engineering (ICSE)},
  pages={27--37},
  year={2017},
  organization={IEEE}
}

@inproceedings{blasi2018translating,
  title={Translating code comments to procedure specifications},
  author={Blasi, Arianna and Goffi, Alberto and Kuznetsov, Konstantin and Gorla, Alessandra and Ernst, Michael D and Pezz{\`e}, Mauro and Castellanos, Sergio Delgado},
  booktitle={Proceedings of the 27th ACM SIGSOFT international symposium on software testing and analysis},
  pages={242--253},
  year={2018}
}

@article{strom2012typestate,
  title={Typestate: A programming language concept for enhancing software reliability},
  author={Strom, Robert E and Yemini, Shaula},
  journal={IEEE transactions on software engineering},
  number={1},
  pages={157--171},
  year={2012},
  publisher={IEEE}
}

@article{xu2000safety,
  title={Safety checking of machine code},
  author={Xu, Zhichen and Miller, Barton P and Reps, Thomas},
  journal={ACM SIGPLAN Notices},
  volume={35},
  number={5},
  pages={70--82},
  year={2000},
  publisher={ACM New York, NY, USA}
}

@article{bierhoff2007modular,
  title={Modular typestate checking of aliased objects},
  author={Bierhoff, Kevin and Aldrich, Jonathan},
  journal={ACM SIGPLAN Notices},
  volume={42},
  number={10},
  pages={301--320},
  year={2007},
  publisher={ACM New York, NY, USA}
}

@inproceedings{yamaguchi2014modeling,
  title={Modeling and discovering vulnerabilities with code property graphs},
  author={Yamaguchi, Fabian and Golde, Nico and Arp, Daniel and Rieck, Konrad},
  booktitle={2014 IEEE symposium on security and privacy},
  pages={590--604},
  year={2014},
  organization={IEEE}
}

@inproceedings{yamaguchi2015automatic,
  title={Automatic inference of search patterns for taint-style vulnerabilities},
  author={Yamaguchi, Fabian and Maier, Alwin and Gascon, Hugo and Rieck, Konrad},
  booktitle={2015 IEEE symposium on security and privacy},
  pages={797--812},
  year={2015},
  organization={IEEE}
}

@inproceedings{lekssays2025llmxcpg,
  title={$\{$LLMxCPG$\}$:$\{$Context-Aware$\}$ vulnerability detection through code property $\{$Graph-Guided$\}$ large language models},
  author={Lekssays, Ahmed and Mouhcine, Hamza and Tran, Khang and Yu, Ting and Khalil, Issa},
  booktitle={34th USENIX Security Symposium (USENIX Security 25)},
  pages={489--507},
  year={2025}
}

@article{ruan2024specrover,
  title={Specrover: Code intent extraction via llms},
  author={Ruan, Haifeng and Zhang, Yuntong and Roychoudhury, Abhik},
  journal={arXiv preprint arXiv:2408.02232},
  year={2024}
}

@article{landi1992undecidability,
  title={Undecidability of static analysis},
  author={Landi, William},
  journal={ACM Letters on Programming Languages and Systems (LOPLAS)},
  volume={1},
  number={4},
  pages={323--337},
  year={1992},
  publisher={ACM New York, NY, USA}
}

@misc{io_safety,
  title        = "IO Safety",
  author       = "Rust Community",
  url          = "https://rust-lang.github.io/rfcs/3128-io-safety.html",
  year         = "2026",
  note         = "[Accessed: 2026-02-01]"
}

@misc{rust_for_linux,
  author = {{Rust-for-Linux Contributors}},
  title = {Rust for {Linux}},
  year = {2026},
  url = {https://rust-for-linux.com/},
  note = {Accessed: 2026-04-22}
}

@misc{bevy_engine,
  author = {{Bevy Contributors}},
  title = {Bevy: A Data-Driven Game Engine Built in {Rust}},
  year = {2026},
  url = {https://bevyengine.org/},
  note = {Accessed: 2026-04-22}
}

@misc{rust_rfc_3842,
  author = {{Rust Community}}, 
  title = {RFC: Safety Tags for Unsafe Code},
  year = {2026},
  howpublished = {\url{https://github.com/rust-lang/rfcs/pull/3842}},
  note = {GitHub Pull Request \#3842. Accessed: 2026-04-23}
}


\end{document}